\def\makeSM{1}
\documentclass[twocolumn,prr,longbibliography,superscriptaddress,floatfix,footinbib]{revtex4-1}

\usepackage[dvipsnames]{xcolor}
\usepackage{graphicx}
\usepackage{amssymb}
\usepackage{amsmath}
\usepackage{amsfonts}
\usepackage{mathtools}
\colorlet{Mycolor1}{Plum}
\usepackage{tikz}
\usepackage[normalem]{ulem}

\newcommand{\ket}[1]{\ensuremath{\left| #1 \right\rangle}}
\newcommand{\bra}[1]{\ensuremath{\left\langle #1 \right|}}
\newcommand{\sand}[2]{\left\langle #1| #2\right\rangle}

\usepackage{bm}

\newcommand{\cd}{\hat{c}^{\dagger}}
\newcommand{\hd}{\hat{h}^{\dagger}}
\newcommand{\co}{\hat{c}}
\newcommand{\h}{\hat{h}}
\newcommand{\VWF}{|\psi(a)\rangle}
\newcommand{\VWFbra}{\langle \psi(a)|}

\newcommand{\Emin}{E_{min}}

\newcommand{\ESYK}{E_{SYK}}
\newcommand{\FSYK}{F_{SYK}}
\newcommand{\EE}{S^{(2)}}
\newcommand{\vac}{|\tilde{0}\rangle}
\newcommand{\vacbra}{\langle\tilde{0}|}
\newcommand{\rthermL}{\rho_{L,pSYK}}
\newcommand{\rthermR}{\rho_{R,pSYK}}
\newcommand{\Etherm}{E_{pSYK}}
\newcommand{\Hres}{H_{pSYK}}
\newcommand{\Top}{\hat{T}}
\newcommand{\Tdop}{\hat{T}^\dagger}
\newcommand{\dtau}{\mathrm{d}\tau}

\newcommand{\avg}[1]{\left< #1 \right>}
\newcommand{\eqn}[1]{eqn.~\ref{#1}}

\newcommand{\fig}[1]{fig.~\ref{#1}}
\newcommand{\Fig}[1]{Fig.~\ref{#1}}
\newcommand{\Refin}[1]{Ref.~\onlinecite{#1}}

\newcommand{\eqns}[2]{eqns.~\ref{#1}, \ref{#2}}

\newcommand{\be}{\begin{equation}}
\newcommand{\ee}{\end{equation}}

\newcommand{\bea}{\begin{equation}\begin{aligned}}
\newcommand{\eea}{\end{aligned}\end{equation}}

\newcommand{\Tr}{\mathrm{Tr}}
\newcommand{\norm}{\mathcal{N}}
\newcommand{\cb}{\bar{c}}
\newcommand{\hb}{\bar{h}}

\newcommand{\rR}{\mathcal{R}}

\renewcommand{\ln}{\log}
\usepackage[colorlinks=true,
            linkcolor=red,
            citecolor=blue,
            urlcolor=blue]{hyperref}

\newcommand{\PRB}[1]{}

\newcommand{\appcite}[1]{SM.~\ref{#1}}

\newcommand{\Arijit}[1]{}

\newcommand{\q}{{(q/2)}}
\newcommand{\qo}{2}
\newcommand{\qt}{4}
\newcommand{\qf}{{q\over2}}
\newcommand{\tq}{q}
\newcommand{\Jij}{J_{i_1\cdots i_\q;j_1\cdots j_\q}}

\newcommand{\coupled}{two-flavor~}

\newcommand{\seccite}[1]{Section~\ref{#1}}
\def\makeSM{1}
\ifdefined\makeSM{}
\newcommand{\SMcite}[1]{Appendix~\ref{#1}}
\else
\newcommand{\SMcite}[1]{\cite{Note3}}
\fi
\begin{document}

\def\titlename{Variational wavefunctions for Sachdev-Ye-Kitaev models}
\title{\titlename}
\def\authornames{Arijit Haldar, Omid Tavakol, Thomas Scaffidi}
\author{Arijit Haldar}
\email{arijit.haldar@utoronto.ca}
\author{Omid Tavakol}
\author{Thomas Scaffidi}

\def\affiliations{Department of Physics, University of Toronto, 60 St. George Street, Toronto, Ontario, M5S 1A7, Canada}
\affiliation{\affiliations}

\begin{abstract}
Given a class of $q$-local Hamiltonians, is it possible to find a simple variational state whose energy is a finite fraction of the ground state energy in the thermodynamic limit?
Whereas product states often provide an affirmative answer in the case of bosonic (or qubit) models, we show that Gaussian states fail dramatically in the fermionic case, like for the Sachdev-Ye-Kitaev (SYK) models.
This prompts us to propose a new class of wavefunctions for SYK models inspired by the variational coupled cluster algorithm.
We introduce a static (``0+0D") large-$N$ field theory to study the energy, two-point correlators, and entanglement properties of these states.
Most importantly, we demonstrate a finite disorder-averaged approximation ratio of $r \approx 0.62$ between the variational and ground state energy of SYK for $q=4$. 
Moreover, the variational states provide an exact description of spontaneous symmetry breaking in a related two-flavor SYK model.

\end{abstract}

\maketitle
\section{Introduction}
Variational wavefunctions are at the heart of our understanding of a variety of condensed matter systems like quantum Hall systems ~\cite{Laughlin}, superconductors~\cite{BCS}, and correlated metals~\cite{Gutzwiller}.
These wavefunctions provide an intuitive description of these phases, and are often useful for numerics.
Working with pure states also makes it possible to study entanglement, a property which has been crucial to characterize exotic phases of matter~\cite{Entanglement}. Further, with the advent of quantum simulators~\cite{Feynman,Lloyd1073}, and in particular of hybrid quantum-classical variational algorithms~\cite{McClean_2016,IBM,HF_Google}, it is desirable to find preparable states that can reach low energy regimes of strongly correlated Hamiltonians. 
 
 A related topic of recent interest is Hamiltonian complexity~\cite{TCS-066}, which studies the computational complexity of approximating the ground state of certain classes of Hamiltonians.
 These problems belong to the quantum Merlin-Arthur (QMA) class since a verifier can check a solution (i.e. a quantum state) efficiently on a quantum computer by measuring its energy~\cite{kitaev2002classical,10.1007/978-3-540-30538-5_31,watrous2008quantum}.
 Whereas approximating the ground state energy within a small additive error was shown to be QMA-complete for a wide range of Hamiltonians, the complexity of approximating the ground state energy density within finite relative error is still undecided, and is closely related to the quantum PCP~\cite{doi:10.1137/110842272,aharonov2013guest,qPCP,8104078,nirkhe_et_al:LIPIcs:2018:9095} and NLTS conjectures \cite{freedman2013quantum}.
Proving these conjectures would, roughly speaking, require finding classes of Hamiltonians for which not only the ground state but all states below a finite energy density are impossible to reach with a simple ansatz.
 
Given a class of traceless Hamiltonians and a class of ansatz wavefunctions, one can define a figure of merit called approximation ratio, given by $r_\psi \equiv E_{\psi}/E_{GS}$, where $E_{GS}$ is the energy of the ground state, and $E_{\psi} = \min_{\psi} \bra{\psi}H\ket{\psi}$, where $\psi$ belongs to the class of ansatz wavefunctions.
For non-trivial Hamiltonians, simple wavefunctions (e.g. product states) are of course not expected to reach an approximation ratio very close to 1. The question we aim to answer instead is whether they can at least achieve $r_{\psi}>0$ in the thermodynamic limit.
Remarkably, the answer can be shown to be affirmative for a variety of bosonic (or qubit) models~\cite{lieb1973,doi:10.1137/110842272,qPCP,PhysRevA.87.062334,doi:10.1063/1.5085428,gharibian2019almost}.
For example, for traceless 2-local qubit Hamiltonians of the type 
\begin{align}
H = \sum_{i,j=1}^N \sum_{\mu,\nu=1}^3 J_{i,j}^{\mu,\nu} \sigma_i^\mu \sigma_j^\nu,
\end{align}
where $\sigma_j^\nu$ are Pauli matrices,
Lieb showed that the approximation ratio of product states has a lower bound: $r_\text{prod} \geq 1/9$~\cite{lieb1973,doi:10.1063/1.5085428}.

Our work is motivated by the following question: can similar results be obtained for $q$-local \emph{fermionic} Hamiltonians~\cite{doi:10.1063/1.5085428}?
For fermionic systems, a natural analog of product states are Gaussian states, which include the Slater determinants calculated with Hartree-Fock. 
However, for $q > 2$, we will provide strong evidence that the approximation ratio of Gaussian states goes to 0 in the thermodynamic limit: $r_\text{Gauss} \to 0$ for $N \to \infty$.
This highlights a fundamental difference between the bosonic and fermionic case.
It also motivates the following question: if Gaussian states are not up to the task, is there any other class of tractable wavefunctions that could provide a finite approximation ratio?

Rather than trying to make statements about all problems in the class, we study instances of $q$-local fermionic Hamiltonians that are typical for a natural measure, which enables us to establish relations with the statistical mechanics of disordered quantum systems.
Namely, we will focus on a paradigmatic example of disordered fermionic systems, the Sachdev-Ye-Kitaev (SYK) models \cite{Sachdev1993,KitaevKITP,Maldacena2016}.
The model has become a primary platform for studying non-Fermi liquid regimes \cite{Parcollet1997,Gu2017,BanerjeeAltman2016,SJian2017b,Song2017,Davison2017,Arijit2017,Patel2018,
Chowdhury2018,Haldar2018PRB,Haldar2020arXivRenyi}, quantum many-body chaos and operator complexity\cite{BanerjeeAltman2016,Maldacena2016,Scaffidi2019PRB,Scaffidi2019PRX}, thermalization\cite{Eberlein2017,Haldar2020,Almheiri2019}, and dualities between quantum-field theory and gravity\cite{Sachdev2010,KitaevKITP,Sachdev2015,Maldacena2018lmt,kitaev2018soft,qi2020coupled}. Whereas a lot is known already about thermal ensembles in SYK models, less is known about wavefunctions of typical low-energy states.
In fact, existing work on pure states in SYK models has relied on thermal states in disguise, like thermofield double states\cite{gao2017traversable,maldacena2017diving} and Kourkoulou-Maldacena (KM) states\cite{Kourkoulou2017,zhang2020entanglement}, and thus require computation in a thermal field theory.
We will propose instead a class of variational wavefunctions for which equal time observables can be computed within a static (``0+0D'') field theory.

This paper is organized as follows. In \seccite{sec:Model}, we formally define the $q$-SYK model and construct a variational ansatz for the model. In \seccite{sec:largeN}, we show that the energy and particle density of the ansatz can be evaluated exactly in the large-$N$ limit. In the same section, we compare the analytical predictions for the ansatz with those obtained using the thermal-field theory of the SYK model. We discuss the nature of entanglement for the ansatz in \seccite{sec:Entanglement} by computing the scaling of the second R\'{e}nyi entropy with subsystem size. Finally, we provide a discussion of our findings in \seccite{sec:Discussion}. Additional details about various results are provided in the appendices and referred to in the main text.
\section{Model and ansatz.}\label{sec:Model}
The $q$-SYK model is defined as:
 \begin{align}
	H_{SYK}=g\sum_{
    \mathclap{\substack{1\leq i_{1}<...<i_{\q} \leq N,\\1 \leq j_{1}<...<j_{\q} \leq N}}}
    J_{i_1\cdots i_\q;j_1\cdots j_\q}
\cd_{i_{1}}..\cd_{i_{\q}}\co_{j_1}..\co_{j_\q},
\label{eqn:HSYK}
 \end{align}
 with $i,j \in [1,N]$ and with $g={\q!}/{\sqrt{\q}(\frac{N}{2})^{\qf-\frac{1}{2}}}$.
 The symbols $\cd_i$, $\co_i$ denote fermionic creation, annihilation operators. The couplings $J_{i_1\cdots i_\q;j_1\cdots j_\q}$ are Gaussian random numbers which satisfy appropriate symmetrization conditions \footnote{E.g., when $q=\qt$, we have $J_{i_1,i_2;j_1,j_2}=-J_{i_2,i_1;j_1,j_2}=J_{i_2,i_1;j_2,j_1}$ due to fermion anti-commutation relations, and $J_{i_1,i_2;j_1,j_2}=J_{j_2,j_1;i_2,i_1}^*$ for maintaining hermiticity.}. The variance is represented as $J^2$, and will be set to one except when written explicitly.
This Hamiltonian has an extensive energy bandwidth which is symmetric around zero due to particle-hole symmetry\footnote{Strictly speaking, particle-hole symmetry is only present in the $N \to \infty$ limit, but all of our results are obtained in that limit anyway.}.

The simplest variational wavefunctions for a fermionic model are Gaussian states (which include Slater determinants), and the corresponding optimization procedure is the celebrated Hartee-Fock~\cite{BLINDER20191}. In quantum chemistry, this technique typically recovers 99\% of the electronic energy, and is the basis for a variety of more sophisticated approaches.
By contrast, for SYK models with $q>2$, an elementary calculation  
(see\ifdefined\makeSM{} \SMcite{sec:slater-proof}\fi) shows that the energy bandwidth of Gaussian states (which is centered around zero) scales subextensively with $N$.
In the large-$N$ limit, Gaussian states therefore only reach a vanishing fraction of the full many-body bandwidth of SYK, and their disorder-averaged approximation ratio goes to zero.
This is a strong indication that the ``worst-case'' approximation ratio of Gaussian states for $q$-local fermionic Hamiltonians with $q>2$ goes to 0 in the large-$N$ limit, in contradiction to the conjecture found in Ref.~\cite{doi:10.1063/1.5085428}.
Intuitively, this happens since minimizing the energy requires optimizing over the value of $\tq$-point correlators, but these correlators are over constrained for a Gaussian state: due to Wick's theorem, all higher-order correlators are simple functions of two-point correlators.

\begin{figure}
    \centering
    \includegraphics[width=0.38\textwidth]{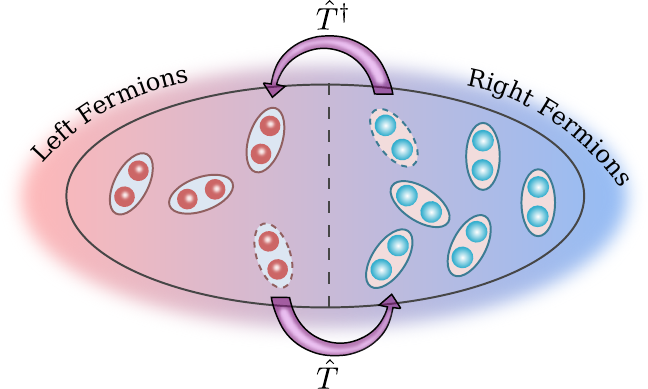}
    
    \caption{Constructing the wavefunction:  The $N$ orbitals are partitioned into left, right subsystems. The operator $\Tdop$  moves $\q$-fermions (with $q=\qt$ in the figure) at a time from the right side to the left side, with the same amplitude $J_{i_1\cdots i_\q;j_1\cdots j_\q}$ as the corresponding term in the Hamiltonian. Starting from a state $\vac$ in which the right-side is filled with fermions and the left is empty, the variational wavefunction is constructed by repeated applications of $\Tdop$.
    }
    \label{fig:pSYK}
\end{figure}

Since Hartee-Fock does not produce any useful result, we take a different approach: let us look for a subset of terms in $H$ which commute with each other, and for which the energy can be minimized easily. 
The selected subset of terms should be extensive in order for the state to have a finite approximation ratio, i.e. it should contain a number of terms which scales as $N^{\tq}$.
We propose to construct such a set by partitioning the system into two subsystems (see \fig{fig:pSYK}), with $N_L$ sites on the left and $N_R=N-N_L$ sites on the right, and by keeping only terms with creation operators on the left side, and annihilation operators on the right side:
\begin{align}\label{eqn:clusop}
\Tdop=g \sum_{
    \mathclap{\substack{i_{1}<...<i_{\q} \in L,\\j_{1}<...<j_{\q} \in R}}} J_{i_1\cdots i_\q;j_1\cdots j_\q}\cd_{i_{1}}...\cd_{i_{\q}} \co_{j_{1}}...\co_{j_{\q}}
\end{align}
where $L=[1,\dots,N_L]$ and $R=[N_L+1,\dots,N]$.
The parameter $p=N_R/N_L$ can be tuned at will, but we will focus on $p=1$ for now. It will be useful to define the partitioned-SYK Hamiltonian, 
\begin{align}\label{eq:HpSYK-def}
\Hres = \hat{T} +\hat{T}^\dagger,
\end{align}
which contains an extensive subset of the terms of $H_{SYK}$, and which is an example of the systems studied in \Refin{Haldar2018PRB}.

Using this notation, the ansatz wavefunction is defined as
\begin{equation}\label{eqn:VWFCE}
     \VWF=\frac{1}{\sqrt{\norm}}\exp(-a \Tdop)\vac,
\end{equation}
where $\vac$ is the state for which all states on the right (resp. left) are full (resp. empty), $a$ is a real variational parameter, and where the normalization is given by $\norm(a)=\vacbra\exp(-a \Top)\exp(-a \Tdop)\vac$.
The intuition behind this state is as follows: starting from a state that is empty on the left and fully occupied on the right, we create a population of particles on the left and holes on the right by applying the corresponding terms from the Hamiltonian. 

Interestingly, this wavefunction belongs to the class of variational coupled cluster (VCC) states developed for quantum chemistry~\cite{vCC1,vCC2,vCC3,Bastianello2016NPB}.
This algorithm has the advantage of being variational (as opposed to regular coupled cluster~\cite{coester1960short,vcivzek1966correlation}), but is usually limited to a very small number of orbitals due to the factorial complexity of the method.
By contrast, we were able to perform VCC directly in the large-$N$ limit for a class of SYK models.

The disorder-averaged energy density for the state is given by $E(a)=\frac1{N} \overline{\VWFbra H_{SYK}\VWF}$ and can be calculated (see \SMcite{SM:large-N-ana-WF} for details) using
\begin{align}\label{eqn:EaCCE}
    E(a)&=\frac{1}{N}\overline{\VWFbra\Hres\VWF}=
    -\frac{1}{N} \frac{\partial \overline{\ln(\norm)}}{\partial a},
\end{align}
where we used the fact that the expectation value of the terms which are present in $H_{SYK}$ but not in $\Hres$ vanishes after disorder averaging.

\begin{figure}
    \centering
\includegraphics[scale=1.1]{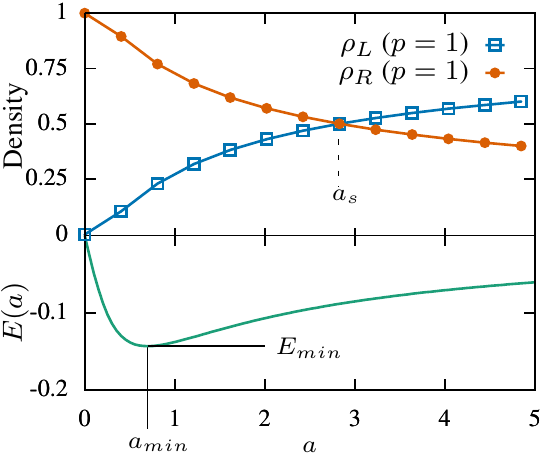} 
    \caption{(Top) Particles densities on the left and right sides ($\rho_L$ and $\rho_R$). The densities are equal at $a_s$. (Bottom) Variational energy, with a minimum at $a_{min}$.}
\label{fig:obsv-vs-a}
\end{figure}

\section{Large-$N$ theory}\label{sec:largeN}
To enable the computation of $\overline{\ln(\norm)}$, we introduce a field-theoretic approach similar to the fermionic path integral (see \SMcite{SM:large-N-ana-WF} for details). First, we perform a particle-hole transformation on the right side, whereby $\co_{i \in R} = \hd_{i \in R}$ and $\cd_{i \in R} = \h_{i \in R}$.
We then define the fermionic-coherent states $|c_{i \in L}\rangle$, $|h_{i \in R}\rangle$ for left and right, characterized by the Grassmann numbers $c_i$, $\cb_i$ and $h_i$, $\hb_i$, respectively, such that $\langle c_i|\cd_i=\langle c_i|\cb_i$, $\langle h_i|\hd_i=\langle h_i|\hb_i$. 
The disorder averaging is implemented using the replica-trick $\overline{\ln(\norm)}=\lim_{\rR\to 0}[\overline{\norm^\rR}-1]/\rR$. 
This results in a ``static'' action involving Grassmann numbers $c_i$, $h_i$ with no imaginary time dynamics. Introducing the static Green's functions 
\begin{align}
	G_c &=-N_L^{-1}\sum_{i\in L}\langle c_i\cb_i\rangle;&
	G_h &=-N_R^{-1}\sum_{i\in R}\langle h_i\hb_i\rangle,
\end{align}
 along with the self-energies $\Sigma_c$, $\Sigma_h$, into the action, allows us to integrate the fermions $c_i$, $h_i$. 
 The particle densities in the left and right subsystems are simply given by $\rho_L=N_L^{-1}\sum_{i\in L}\langle\cd_i \co_i\rangle=1+G_c$ and $\rho_R=N_R^{-1}\sum_{i\in R}\langle\h_i\hd_i\rangle =-G_h$ respectively. For $p=1$, particle conservation implies $\rho_L+\rho_R=1$, and thus $G_c=G_h\equiv G$ and $\Sigma_c=\Sigma_h=\Sigma$.
At the saddle point, one finds
 \begin{align}\label{eqn:SP-cond}
 -G^{-1}&=1+\Sigma \notag\\
\Sigma&=-a^{2}J^2G^{\tq-1},
 \end{align}
 which are polynomial equations for $G(a)$ and $\Sigma(a)$ that can easily be solved numerically.
These relations derive from the generating function $\overline{\ln({\norm})}$, which takes the form
\begin{equation}\label{eqn:lnNSP}
    -\overline{\ln(\norm)}=-N\left[ \ln(1+\Sigma)+\Sigma G +\frac{a^2 J^2}{\tq} G^{\tq}\right],
\end{equation}
at the large-$N$ saddle-point.
Interestingly, this generating functional can be interpreted as a static limit of the free-energy for the SYK model \cite{Maldacena2016,BanerjeeAltman2016},  given by:
\begin{align}\label{eqn:FSYK}
\FSYK=&-N\left[T\ln\det(\partial_\tau+\Sigma)+\int d\tau \Sigma(\tau)G(\beta-\tau)\right.\notag\\
&\left.+(J^2/\tq)\int d\tau G(\tau)^\q G(\beta-\tau)^\q\right],
\end{align}
where $\tau\in[0,\beta]$ denotes the imaginary-time variable and $\beta=T^{-1}$ is the inverse temperature. 
Indeed, if the imaginary time dynamics is eliminated by substituting $ \partial_\tau\to1$, $G(\tau)\to G$ and $\Sigma(\tau)\to\Sigma$, an expression similar to $-\overline{\ln(\norm)}$ in \eqn{eqn:lnNSP} is recovered.

\def\higherqs{}
\ifdefined\higherqs
\begin{figure}
    \centering
    \includegraphics[scale=1.1]{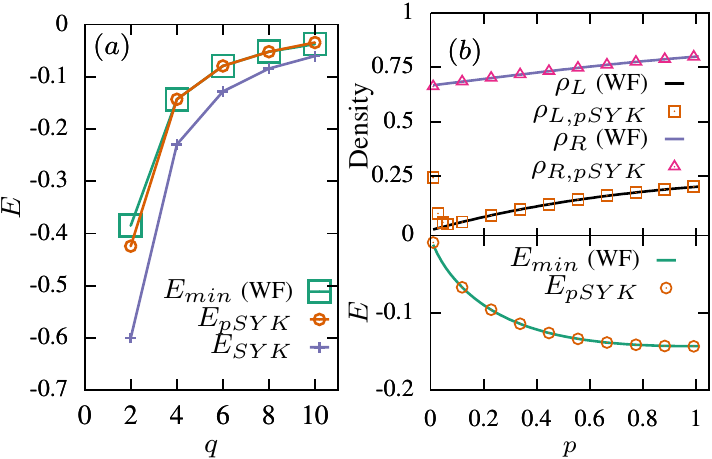} 
    \caption{(a) Comparison between variational energy $\Emin$ and exact ground states energies of $H_{SYK}$ and $H_{pSYK}$, for $p=1$ and varying $q$. We find $\Emin$ and $E_{pSYK}$ to be equal within our numerical accuracy for $q \geq 4$.
    (b) Comparison between variational wavefunction (lines) and exact ground state of $H_{pSYK}$ (symbols), when $q=4$, for the energy and the particle densities in the left and right subsystems. 
    }
    \label{fig:therm-WF-compare}
\end{figure}
\else
\begin{figure}[t!]
    \centering
\includegraphics[scale=1.1]{large-N_vs_WF.pdf} 
    \caption{Comparison of ground-state energy ($\Etherm$, circles), particle density on the left ($\rthermL$, squares) and right ($\rthermR$, triangles) sides obtained from exact thermal-field theory by varying the site-ratio $p=N_R/N_L$, with the predictions obtained from the variational WF ansatz (lines) by minimizing the energy $E(a)$ w.r.t $a$.}
    \label{fig:therm-WF-compare}
\end{figure}
\fi

The energy density $E(a)$ is calculated using \eqn{eqn:EaCCE} to give
\begin{align}
E(a)=- \frac{2}q a J^2 G^{\tq},
\end{align}
where $G$ is obtained by solving the saddle point equations (see \fig{fig:obsv-vs-a}).
The most important point is that $E(a)$ does not decay with $N$, which means the variational states have an extensive bandwidth, and thus a finite approximation ratio in the large-$N$ limit.

The variational energy has a single minimum as a function of $a$ (see \fig{fig:obsv-vs-a} bottom), with the following properties:
\begin{align}\label{eqn:Emin-formula}
a_{min} &= \frac1{J} \frac{(q+1)^{\frac{q-1}{2}}}{q^{\frac{q}{2}}} \\
E_{min} &= - J \frac{2}{q} \frac{q^{\frac{q}2}}{(q+1)^{\frac{q+1}{2}}}.
\end{align}
We can now compare $\Emin$ with the energy density of the ground-state of the SYK model ($\ESYK$). 
The latter can be obtained by taking the zero temperature limit of $\FSYK$ (see \eqn{eqn:FSYK}).
We give a comparison as a function of $q$ in Fig.~\ref{fig:therm-WF-compare}(a).
For example, for $q=4$, we find $\Emin=8/25 \sqrt{5} \approx -0.143$ and $\ESYK\approx-0.2295$.
Since we expect both $\Emin$ and $\ESYK$ to be self-averaging, we define the disorder-averaged approximation ratio as $r_{\psi}=\Emin/\ESYK$.
We thus find $r_{\psi}\approx 0.62$ for $q=4$.
To put things into perspective, we have calculated that $\Emin$ has the same energy density as the thermal ensemble of the SYK model at temperature $T/J\approx 0.455$.

A peculiarity of $\ket{\psi(a)}$ is that the particle densities on the left and right depend on $a$, and are only equal for $a=a_s=2^{\q-1/2}$ (see \fig{fig:obsv-vs-a})\footnote{In usual applications of coupled cluster theory, the partitioning of the system is decided by the Hartee-Fock method, which separates orbitals that are occupied in the Hartree-Fock state from the others. In that setting, the left-right asymmetry is natural, and measures how many particle-hole excitations from the reference Hartree-Fock state are created in order to accommodate the interaction terms in the Hamiltonian. By contrast, in our case the partitioning between left and right is artificial and the fact that $|\psi(a_{min})$\unexpanded{$\rangle$} is unbalanced is an artefact of the technique that should not be present in the true eigenstates of $H_{SYK}$. However, when studying $\Hres$, this balance is actually physical  and is a manifestation of spontaneous symmetry breaking.}.
Since $a_s \neq a_{min}$, the variational state with the lowest energy has an asymmetric particle density between left and right, in contrast to the ground state of the original SYK model for which all orbitals are at half-filling.
This discrepancy arises from the fact that our construction aims at minimizing $\Hres$, which contains only a subset of the terms in $H_{SYK}$, and creates an artificial distinction between the two subsystems.
The Hamiltonian $H_{pSYK}$ is actually interesting in its own right as it can be understood as an example of \coupled SYK models, in which two SYK quantum dots are coupled by $q$-body interactions, as studied in \Refin{Haldar2018PRB}.
For $q \geq 4$, $H_{pSYK}$ was shown to have a low temperature phase which exhibits phase separation: one subsystem (say the one on the left) has density $1/(q+1)$, and the other one has density $q/(q+1)$.
This phase has a gap to single-particle excitations, and spontaneously breaks particle-hole symmetry and left-right interchange symmetry, but conserves their product.

\begin{figure}
    \centering
\includegraphics[scale=1.1]{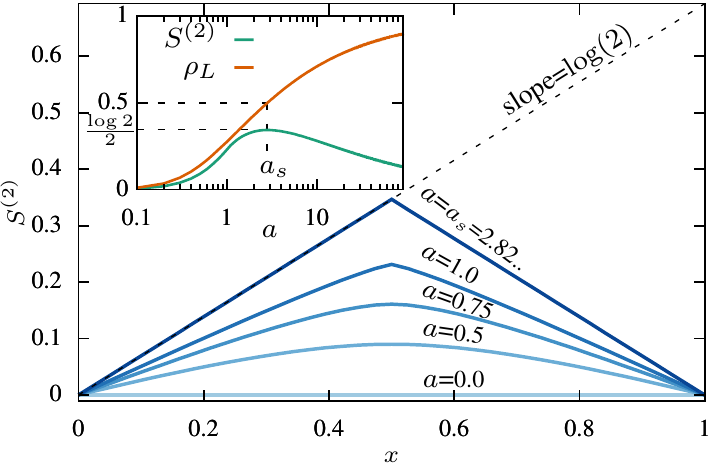}
     \caption{Second R\'{e}nyi entropy $\EE$ of the state $\VWF$ as a function of partition size $x$ for multiple values of $a$, and for $p=1$ and $q=4$. 
     Inset: $\EE(x=0.5)$ and $\rho_L$ as a function of $a$. The state is maximally entangled at the left-right symmetric value of $a=a_{s}$. The entanglement decays monotonically with $a$ beyond that value.}
    \label{fig:S2_vs_x}
\end{figure}

Interestingly, $\ket{\psi(a_{min})}$ reproduces this density imbalance perfectly: we find $\rho_{L,\min} = 1-\rho_{R,\min}=1/(q+1)$.
Further, we find $E_{\min}$ to be equal to the ground state energy of $H_{pSYK}$ (which can be obtained in a similar fashion as $E_{SYK}$, see \SMcite{sec:thermal-field-theory-pSYK}) within numerical accuracy for $q \geq 4$.
We checked that this agreement even extends to the asymmetric case of $p=N_R/N_L \neq 1$.
In the context of \coupled SYK models, this ratio gives the relative size of the two dots~\cite{Haldar2018PRB}.
The comparison for $\rho_L$, $\rho_R$, $\Emin$ with the exact values is shown in \fig{fig:therm-WF-compare}\ifdefined\higherqs (b)\fi, for $q=4$.
 The only discrepancy appears as $p \to 0$, which is expected since $H_{pSYK}$ undergoes an additional phase-transition to a gapless phase at $p_c \simeq 0.072$~\cite{Haldar2018PRB}.
 Another discrepancy appears for $q=2$, in which case the variational wavefunction fails to describe the Fermi liquid phase of $H_{pSYK}$ which survives down to $T=0$ (see \SMcite{sec:one-body-p-SYK} for more details).

\section{Entanglement}\label{sec:Entanglement}
The entanglement properties of $\ket{\psi(a)}$ can also be calculated using a recently developed formalism~\cite{Haldar2020arXivRenyi}. Some of the earlier studies on entanglement in the SYK model can be found in \Refin{Fu2016,BalentsPRB2018,GoelJHEP2019,GuPRD2019}. 
We focus on the second R\'{e}nyi entropy, $\EE=-N^{-1} \overline{\ln\Tr[\hat{\rho}_A^2]}$, for a biparition of the system into regions $A$ and $B$, and where $\hat{\rho}_A=\Tr_B\VWF\VWFbra$ is the reduced density matrix.
The partition is parametrized by $x \in [0,1]$, which gives the proportion of orbitals in $A$.
For $x\le 0.5$ we take region A to be entirely comprised of the left-side fermions, while $x>0.5$ also includes a portion ($x-0.5$) of the right-side fermions.
The large-$N$ limit for $\EE$ is obtained using an approach similar to calculating $\overline{\ln(\norm)}$ (see \SMcite{sup:renyi}).

The results for $\EE$ are shown in \Fig{fig:S2_vs_x} for $q=4$. The $x$ dependence of $\EE$ resembles the one obtained for KM states in SYK~\cite{zhang2020entanglement}, with a small-$x$ linear behavior indicative of a volume law of entanglement, and a maximum at $x=0.5$.
Starting from 0 at $a=0$, the entanglement grows until the left-right symmetric point $a=a_s$ is reached, after which it decays monotonically (see \Fig{fig:S2_vs_x} inset).
Remarkably, we find $\EE(x)=\min(x,1-x) \log(2)$
at $a=a_s$, which means $\ket{\psi(a_s)}$ is maximally entangled between the left and right subsystems.

\section{Discussion}\label{sec:Discussion}
In this work, we have highlighted a fundamental difference between bosonic (or qubit) and fermionic $q$-local Hamiltonians, as regards to the complexity of finding wavefunctions with a finite approximation ratio ($E_{\psi}/E_{GS}>0$) in the thermodynamic limit.
We showed that, for a prototypical fermionic model, the SYK model, the bandwidth of Gaussian states scales subextensively with system size, leaving a parametrically large gap between the ground state and Gaussian states.
This raises the question of whether other classes of tractable wavefunctions could (partially) fill this gap.
We took a step in that direction by proposing a wavefunction inspired by the variational coupled cluster algorithm with a disorder-averaged approximation ratio of $r \approx 0.62$.

From a physical perspective, this wavefunction is easily tractable, since it is described by a static large-$N$ field theory for which saddle point equations are simply given by polynomial equations.
It remains however unpractical from a computational point of view since a ``brute-force'' calculation of its properties would have factorial complexity on a classical computer. Further, to the best of our knowledge, there is no efficient algorithm to prepare a VCC state on a quantum computer.
It is therefore desirable to find other classes of wavefunctions with $r>0$ which could efficiently be studied with a classical or quantum computer. Unitary coupled cluster states are particularly promising regarding the latter possibility~\cite{IBM,McClean_2016,o2016scalable,shen2017quantum}, and could be studied by extending the techniques developed here.

Moreover, our approach of focusing on a subset of terms in the SYK Hamiltonian could be transposed to other versions of SYK models with a reduced number of terms, like low-rank SYK~\cite{PhysRevB.101.125112} and sparse SYK~\cite{garcia2020sparse,2020arXiv200802303X}.
More generally, we surmise that large-$N$ techniques and SYK models could prove a useful tool in the search for new variational wavefunctions.

\begin{acknowledgements}
We would like to acknowledge helpful discussions with Ehud Altman, Xiangyu Cao, Matthias Degroote, Tim Hsieh, Bryce Kobrin, Sumilan Banerjee and Arun Paramekanti.
This research was enabled in part by support provided by Compute Canada (\url{www.computecanada.ca}).
We acknowledge the support of the Natural Sciences and Engineering Research Council of Canada (NSERC), in particular the Discovery Grant [RGPIN-2020-05842], the Accelerator Supplement [RGPAS-2020-00060], and the Discovery Launch Supplement [DGECR-2020-00222]. TS contributed to this work
prior to joining Amazon.
\end{acknowledgements}

\ifdefined\makeSM{}
\appendix

\section{Subextensive scaling of energy for Gaussian states }\label{sec:slater-proof}
In this section we show that the energy bandwidth of Gaussian states is subextensive for the SYK model with $q >2$. We start with a derivation specific to $q=4$, and then treat the more general case by mapping it to a classical spin glass model.

\textit{Case of $q=4$.}---
We use the Majorana version of SYK for convenience, written as
 \begin{align}
	H =-\frac{1}{N^{3/2}}\sum_{
    \mathclap{\substack{ijkl}}}J_{ijkl}
\gamma_{i}\gamma_{j}\gamma_{k} \gamma_{l},
\end{align}
 where $\gamma_{i}$ represent the Majorana fermions with $\{ \gamma_i, \gamma_j \} = \delta_{ij}$. Using Wick's theorem and the permutation properties of $J_{ijkl}$, the expectation value of the Hamiltonian for an arbitrary Gaussian state can be written as
 \begin{align}
     \avg{H} = &-\frac{3}{N^{3/2}}\sum_{
    \mathclap{\substack{ijkl}}}J_{ijkl} \avg{\gamma_i\gamma_j}\avg{\gamma_k\gamma_l} \notag\\
    =& -12 \frac{1}{N^{3/2}}\sum_{
    \mathclap{\substack{i<j,k<l}}}J_{ijkl} \avg{\gamma_i\gamma_j}\avg{\gamma_k\gamma_l}.
 \end{align}
 Interpreting $J_{i<j,k<l}$ as a real symmetric matrix and $L_{i<j} \equiv i \avg{\gamma_i\gamma_j} $ as a vector, we go to the eigenbasis of $J$, leading to
    \begin{align}
    \avg{H} = 12 \frac{1}{N^{3/2}}\sum_{
    \mathclap{\substack{i<j,k<l}}}J_{ijkl} L_{ij} L_{kl} = 12 
        \frac{1}{N^{3/2}}\sum_{
    \mathclap{\substack{\mu}}}\lambda_{\mu}L_{\mu}^2
    \end{align}
    where $\lambda_\mu$ are the eigenvalues of $J_{i<j,k<l}$, and $L_\mu$ are its eigenvectors.
    Minimizing $ \avg{H}$ now amounts to minimizing this quadratic form, but with an extensive number of constraints on the values of $L_\mu$ in order for them to be consistent with a Gaussian state.
    In order to obtain a non-trivial bound on $\avg{H}$, it is sufficient to take into account the simplest of such constraints, which sets the norm of the vector $L$: 
    \begin{align}
    \sum_\mu L_{\mu}^2 = \sum_{i<j}L^{2}_{ij}= N/8.
    \end{align}
    Minimizing the quadratic form under this single constraint is straightforward, and leads to the following bound:
    \begin{align}
    \avg{H} \geq 12 
        \frac{1}{N^{3/2}} \frac{N}8 \lambda_\text{min}
        \label{eq:finalboundS1}
    \end{align}
    where $\lambda_\text{min}$ is the smallest eigenvalue of $J$.

    We now need to find the scaling of $\lambda_\text{min}$.
    In the large $N$ limit, we expect the matrix $J_{i<j,k<l}$ to behave as a random matrix of dimension $O(N^2)\times O(N^2)$, and thus to have a semi-circle distribution of eigenvalues with radius $O(N)$ (this was verified numerically for $N$ up to 200). We therefore expect $\lambda_\text{min}$ to be a negative number of order $N$. From Eq.~\ref{eq:finalboundS1}, this means that the bandwidth of Gaussian states scales at most like $\sqrt{N}$ (whereas the full bandwidth scales like $N$ since it is extensive).

\renewcommand{\baselinestretch}{1.05}

\textit{General case.}---
We can show that the above sub-extensive scaling also holds when $q>4$ by mapping the problem to the $p$-spin spherical spin glass model~\cite{pspin}. To do this, we start from the following Hamiltonian:
\bea
H = \frac{i^{q/2}}{N^{(q-1)/2}} \sum_{i_1 \dots i_q} J_{i_1 \dots i_q} \gamma_{i_1} \dots \gamma_{i_q}.
\eea
We compute the expectation value for a Gaussian state in a similar way as above, leading to:
\bea
\avg{H} =& \frac{i^{q/2}}{N^{(q-1)/2}} \frac{q!}{(q/2)!}\notag\\
&\sum_{ \mathclap{(i_1<i_2),(i_3<i_4),\dots, (i_{q-1}<i_q)}} J_{i_1,\dots,i_q}  \avg{\gamma_{i_{1}}\gamma_{i_2}} \dots \avg{\gamma_{i_{q-1}}\gamma_{i_q}}.
\eea
Denoting $a_1=(i_1<i_2)$ and similarly for the other indices, we rewrite the expectation value as
\bea
\avg{H} =  \frac{1}{N^{(q-1)/2}} \frac{q!}{(q/2)!} \sum_{a_1,\dots,a_{q/2}} J_{a_1\dots a_{q/2}} L_{a_1} \dots L_{a_{q/2}},
\label{eq:pspin-0}
\eea
where $L_{a}$ is again understood as a $N(N-1)/2$-dimensional vector. Even though there exists a large number of constraints on the vector $L$, we find again that it is sufficient to impose the simplest one ($\sum_a L_a^2 = N/8$) to obtain a non-trivial bound. This will provide the spherical constraint for the mapping to the spherical $p$-spin model.

The $p$-spin spherical model is defined as~\cite{pspin}
\bea
H_\text{p-spin} = \frac1{M^{(p-1)/2}} \sum_{a_1 \dots a_p=1}^M J_{a_1 \dots a_p} s_{a_1} \dots s_{a_p},
\eea
with $a \in 1,\dots,M$ and $\sum_a s_a^2 = M$, and where $J_{a_1 \dots a_p}$ are Gaussian-distributed random numbers. This model is extensive: its bandwidth scales like $M$, the number of classical spins.

We can now make the following identifications:
\begin{align}
p&=q/2 \\
M&=\frac{N(N-1)}2 \\
s_a &=  2 \sqrt{N-1} L_a 
\end{align}
in order to relate the two models. This finally leads to
\bea
\avg{H} \geq \frac{1}{N^{(q-1)/2}} \frac{q!}{(q/2)!} \frac1{(2 \sqrt{N-1})^{q/2}} M^{(p-1)/2} E_\text{p-spin},
\label{eq:pspin}
\eea
where $E_\text{p-spin}$ is the ground state energy of an instance of the spherical $p$-spin model for which the couplings $J_{a_1,\dots,a_p}$ are given by the corresponding $J_{(i_1<i_2),\dots,(i_{q-1},i_q)}$ of the SYK Hamiltonian.
We now make the assumption that these instances of the $p$-spin spherical model are typical, or in other words that the correlations present in $J_{(i_1<i_2),\dots,(i_{q-1},i_q)}$ due to permutation symmetries can be neglected.
If that is the case, we can use the fact that the spherical $p$-spin model is extensive to deduce that $E_\text{p-spin} $ scales like $M \sim O(N^2)$.
By using this relation, the right-hand side of Eq.~\ref{eq:pspin} can be shown to scale like $N^{\frac32 - \frac{q}4}$.
The bandwidth of Gaussian states therefore scales at most like $N^{\frac32 - \frac{q}4}$, which is subextensive for $q>2$.
Setting $q=2$, we find a Gaussian state bandwidth which is extensive, as expected since in that case the ground state is a Gaussian state. For $q=4$, we find $\sqrt{N}$ as previously shown. For larger $q$, the Gaussian states' bandwidth gets narrower and narrower.

\section{Large-$N$ analysis of the variational wavefunction}
\label{SM:large-N-ana-WF}
In this section, we discuss the details pertaining to the computation of $\overline{\log(\mathcal{N})}$ (see \eqn{eqn:lnNSP}) in the large-$N$ limit. As stated in the main text, the said quantity works as a generating functional for computing observables and correlation functions for the variational wavefunction. Since calculating the disorder average of the $\ln$-term directly is hard, we use the replica trick to represent the  term as
\begin{equation}
    \overline{\ln(\norm)}=\lim_{\rR\to 0}\frac{\overline{\mathcal{N}^{\rR}-1}}{\rR},
    \label{eqn:replica-trick-app}
\end{equation}
where $\rR$ denotes the number of replicas. The normalization $\norm(a)=\vacbra\exp(-a \Top)\exp(-a \Tdop)\vac$ (see \eqn{eqn:clusop} for the definition of $\Top$) can be written as an integral over the fermionic-coherent states $|c_{i\in L}\rangle$, $|h_{j\in R}\rangle$, representing the particles and holes, such that 
\begin{widetext}
\begin{align}
    \norm^\rR=&\left(\int\mathcal{D}[c,h]
    \vacbra\exp(-a \Top)\ket{c_i,h_j}\bra{c_i,h_i}\exp(-a \Tdop)\vac\right)^\rR\notag\\
    =&\int\mathcal{D}[c,h]\exp\left[\sum_{r=1}^{\rR}{\left(-\sum_{i\in L}\cb_{i,r}c_{i,r}-\sum_{j\in R}\hb_{j,r}h_{j,r}\right)}\right.\notag\\
    & +\left.\sum_r\ \ \ (-a)g\sum_{\mathclap{\substack{i_{1}<...<i_{(q/2)} \in L,\\j_{1}<...<j_{(q/2)} \in R}}}\big( \Jij\cb_{i_1,r}\cdots\cb_{i_{(q/2)},r}\hb_{j_{(q/2)},r}\cdots\hb_{j_1,r}+
    \Jij^{*}h_{j_1,r}\dots c_{i_1,r}\big)\right],
    \label{eqn:nom-pow-R-exp}
\end{align}
\end{widetext}
where the Grassmann-numbers $\cb_{ir}$, $c_{ir}$, $\hb_{ir}$, $h_{ir}$ are indexed by the replica index $r$ and the site-index $i$. Contrary to usual thermal-field theory, the Grassmann-numbers do not require an imaginary-time $\tau$ index since the terms in the cluster-operator $\hat{T}$ commute. Disorder averaging \eqn{eqn:nom-pow-R-exp} over all possible realizations of $\Jij$, gives us
\begin{align}\label{eqn:Nr-disavg}
    \overline{\norm^{R}}=&\int \mathcal{D}[c,h]
    \exp\left[\sum_{r} \left(-\sum_{i\in L}\cb_{ir}c_{ir}-\sum_{j\in R}\hb_{jr}h_{jr}\right)\right.\notag\\
        &+\frac{2a^2J^2}{q(\sqrt{N_L N_R})^{q-1}}\notag\\
        &\left.\sum_{r_1,r_2}\left(\sum_{i\in L}\cb_{i,r_1}c_{i,r_2}\right)^{q/2}
        \left(\sum_{j\in R}\hb_{j,r_1}h_{j,r_2}\right)^{q/2}\right].
\end{align}
To obtain the large-$N$ limit of the above integral, we introduce the static Green's function $G_c$, $G_h$ , and demand that they must satisfy
\begin{align}\label{eqn:s-GF-app}
    G_{c}(r_1,r_2)=&-\frac{1}{N_L}\sum_{i\in L}\langle c_{ir_1}\cb_{ir_2}\rangle\notag\\
    G_{h}(r_1,r_2)=&-\frac{1}{N_R}\sum_{j\in R}\langle h_{ir_1}\hb_{ir_2}\rangle
\end{align}
at the large-$N$ saddle point.
The above constraints can be incorporated into \eqn{eqn:Nr-disavg} using the static self-energies $\Sigma_c$, $\Sigma_h$ such that
\begin{align}
\overline{\norm^\rR}= &\int\mathcal{D}[c,h]\mathcal{D}[G,\Sigma] \notag\\
&\exp\left[\sum_{r_1,r_2,i}{-\cb_{ir_1}(\delta_{r_1,r_2}+\Sigma_c(r_1,r_2))c_{ir_2}}\right.\notag\\
&\left.-\sum_{r_1,r_2,j}\hb_{jr_1}(\delta_{r_1,r_2}+\Sigma_h(r_1,r_2))h_{jr_2}\right]\notag\\
&
\exp\left[\sum_{r_1,r_2}
N_L \Sigma_{c}(r_1,r_2)G_c(r_2,r_1)\right.\notag\\
&\qquad+N_R\Sigma_{h}(r_1,r_2)G_{c}(r_2,r_1)\notag\\
&\left.+\frac{2a^2J^2\sqrt{N_L N_R}}{q} G_{c}(r_1,r_2)^{q/2}G_{h}(r_1,r_2)^{q/2}\right]
\end{align}
where $\Sigma_{c,h}$ act as Lagrange multipliers. We integrate out the fermions from the above to get
\begin{align}\label{eqn:NR-action}
\overline{\norm^\rR}=&\int \mathcal{D}[G,\Sigma] \exp\left[-S[G,\Sigma]\right]\notag\\     S[G,\Sigma]=&-N_L\ln\left(\det(\mathbf{1}+\Sigma_{c})\right)-N_R\ln\left(\det(\mathbf{1}+\Sigma_{h})\right)\notag\\
&-\sum_{r_1,r_2}{\bigg(}
N_L \Sigma_{c}(r_1,r_2)G_c(r_2,r_1)\notag\\
&+N_R\Sigma_{h}(r_1,r_2)G_{h}(r_2,r_1)\notag\\
&\left.+\frac{2a^2J^2\sqrt{N_L N_R}}{q} G_{c}(r_1,r_2)^{q/2}G_{h}(r_1,r_2)^{q/2}\right),
\end{align}
where we have introduced the effective action $S[G,\Sigma]$, and $\mathbf{1}$ represents the identity matrix in the replica-space. We evaluate the integral in \eqn{eqn:NR-action} at the saddle-point for the action $S$. Furthermore, we shall consider a replica-diagonal ansatz for $G_{c,h}$, $\Sigma_{c,h}$, i.e. $G_{c,h}(r_1,r_2)=\delta_{r_1,r_2}G_{c,h}$ and $\Sigma_{c,h}(r_1,r_2)=\delta_{r_1,r_2}\Sigma_{c,h}$. This results in the following simplified form for the effective action
\begin{align}\label{eqn:SP-ch-app}
    S[G,\Sigma]=&-\rR N(1+p)^{-1}{\bigg(}\ln\left(1+\Sigma_{c}\right)+p\ln\left(1+\Sigma_{h}\right)\notag\\
&\left.+\Sigma_{c}G_c+p\Sigma_{h}G_h
+\frac{2a^2J^2\sqrt{p}}{q} G_{c}^{q/2}G_{h}^{q/2}\right),
\end{align}
where we have used the site-ratio $p=N_R/N_L$ and the total number of sites $N=N_R+N_L$. Minimizing the above replica-diagonal action with respect to $G_{c,h}$, $\Sigma_{c,h}$ we get the saddle-point conditions
\begin{align}\label{eqn:SP-cond-app}
    \left(1+\Sigma_{c,d}\right)^{-1}	=&-G_{c,d}\notag\\
\Sigma_{c}	=&-\sqrt{p}a^{2}J^{2}G_{c}^{q/2-1}G_{d}^{q/2}\notag\\
\Sigma_{d}	=&-a^{2}(1/\sqrt{p})J^{2}G_{d}^{q/2-1}G_{c}^{q/2}.
\end{align}
The value for $\overline{\ln(\norm)}$ at the saddle-point is given by
\begin{align}\label{eqn:lnN-app}
    \overline{\ln(\norm)}=&\lim_{\rR\to 0}\frac{\exp^{-S[G,\Sigma]}-1}{\rR}\notag\\
                       =&N(1+p)^{-1}{\Big(}\ln\left(1+\Sigma_{c}\right)+p\ln\left(1+\Sigma_{h}\right)\notag\\
&\left.+\Sigma_{c}G_c+p\Sigma_{h}G_h
+\frac{2a^2J^2\sqrt{p}}{q} G_{c}^{q/2}G_{h}^{q/2}\right).
\end{align}
The expression for $\overline{\ln(\norm)}$ given in \eqn{eqn:lnNSP} of the main-text is then obtained by setting $p=1$ and $G_c=G_h=G$, $\Sigma_c=\Sigma_h=\Sigma$, so that
\begin{equation}\label{eqn:lnNSP-app}
 \overline{\log(\norm)}=N\left[ \log(1+\Sigma)+\Sigma G +\frac{a^2 J^2}{q} G^{q}\right].
\end{equation}
Similarly, the saddle-point conditions in \eqn{eqn:SP-cond-app} take the form
\begin{align}
    -G^{-1}=&1+\Sigma\notag\\
    \Sigma=&-a^2J^2G^{q-1},
\end{align}
as reported in \eqn{eqn:SP-cond} of the main-text.\\

\paragraph*{Energy Density}
Having obtained the saddle-point solutions $G_c$, $G_h$ we can now calculate the energy density for the ansatz with respect to the full SYK Hamiltonian which can be written as $H_{SYK}=\Hres+H_{other}$. The $\Hres$ (defined in \eqn{eq:HpSYK-def}) encodes the scattering of $q/2$-fermions from left(L) to right(R) and vice-versa, whereas $H_{other}$ denotes the other scattering processes not accounted by $\Hres$. It is easy to show that, after disordering averaging over $\Jij$,
\begin{align}
    E(a)&=\frac1{N}\overline{\langle \psi(a)|H_{SYK}|\psi(a)\rangle}\notag\\
    &=\frac1{N}\overline{\langle \psi(a)|\Hres|\psi(a)\rangle}+\frac1{N}\underbrace{\overline{\langle \psi(a)|H_{other}|\psi(a)\rangle}}_{=0}\notag\\
    &=\frac1{N}\overline{\langle \psi(a)|\Hres|\psi(a)\rangle},
\end{align}
since the $\Jij$ in $H_{other}$ will appear odd-number of times and average out to \emph{zero}. We now work with $\langle \psi(a)|\Hres|\psi(a)\rangle$. Since $\Hres = \Top +\Tdop$(see \eqn{eq:HpSYK-def}), we have
\begin{align}
    -\partial_a\log(\norm)&=-\frac{1}{N}\partial_a\log\left(\vacbra\exp(a\Top)\exp(-a \Tdop)\vac\right)\notag\\
    &=\frac{1}{N}\frac{\vacbra\exp(a\Top)(\Top+\Tdop)\exp(-a \Tdop)\vac}{\norm}\notag\\
    &=\langle\psi(a)|\Hres|\psi(a)\rangle,
\end{align}
i.e., $E(a)=\frac{1}{N}\overline{\VWFbra\Hres\VWF}=-(1/N)\partial_a \overline{\ln(\norm)}$ which was reported in \eqn{eqn:EaCCE} of the main text. Proceeding forward, we can calculate $E(a)$ from $G_c$, $G_h$ as shown below
\begin{align}\label{eqn:energy-WF-app}
    E(a)=&\frac1{N}\overline{\langle \psi(a)|\Hres|\psi(a)\rangle}=-\partial_{a}\frac{\overline{\ln\norm}}{N_{L}+N_{R}}\notag\\
    =&-|a|\frac{4J^{2}}{q}\frac{\sqrt{p}}{1+p}G_{c}^{q/2}G_{h}^{q/2},
\end{align}
where we have used \eqn{eqn:lnN-app} to take the derivative.
\paragraph*{Density of particles and holes}
The density of particles, say for the left-side fermions, is obtained by calculating expectation value 
\begin{equation}
\rho_L=\overline{\frac{\vacbra \exp(-a\Top)\cd_{i}\co_{i}\exp(-a\Tdop)
\vac}{\vacbra\exp(-a\Top)\exp(-a\Tdop)\vac}}
\end{equation}
in the large-$N$ limit. Instead of evaluating the above expression directly, we use a chemical-potential-like source term $\mu$, such that
\begin{equation}\label{eqn:rhoL-gen-func-app}
\rho_{L}=\partial_{\mu\to0}\overline{\ln\vacbra\exp(-a\Top)\exp(\mu\cd_{i}\co_{i})\exp(-a\Tdop)\vac}.
\end{equation}
The advantage of using a source-term is that we can repeat the same analysis used for computing the energy earlier (\eqn{eqn:energy-WF-app}) in this case as well. At the end of which we get the following saddle point equations
\begin{align}
\left(1+(1+\mu)\Sigma_{c}\right)^{-1}(1+\mu) & =-G_{c}\notag\\
\left(1+\Sigma_{h}\right)^{-1} & =-G_{h}\notag\\
\Sigma_{c} & =-a^{2}J^{2}G_{c}^{q/2-1}G_{h}^{q/2}\notag\\
\Sigma_{h} & =-a^{2}J^{2}G_{h}^{q/2-1}G_{c}^{q/2},
\end{align}
that give back the saddle-point conditions of \eqn{eqn:SP-cond-app} in the $\mu\to0$ limit. The corresponding replica-diagonal action for the $\overline{\ln(\cdots)}$ term in \eqn{eqn:rhoL-gen-func-app} is found to be
\begin{align}
    S_{\rho_{L}}(\mu)=&-\rR{\Big[}N_L\ln\det\left[1+(1+\mu)\Sigma_{c}\right]\notag\\
    &+N_R\ln\det\left[1+\Sigma_{h}\right]+\frac{2a^2J^2\sqrt{N_L N_R}}{q}G_{c}^{q/2}G_{d}^{q/2}\notag\\
    &+N_L\Sigma_{c}G_{c}+N_R\Sigma_{h}G_{h}{\Big]}.
\end{align}
Using the fact $\overline{\ln\vacbra\exp(-a\Top)\exp(\mu\cd_{i}\co_i)\exp(-a\Tdop)\vac}= S_{\rho_{L}}(\mu)/\rR$, we compute the derivative of $S_{\rho_{L}}(\mu)/\rR$ w.r.t $\mu$ as shown below 
\begin{align*}
&\partial_{\mu\to0}\overline{\ln\vacbra\exp(-a\Top)\exp(\mu\cd_{i}\co_{i})\exp(-a\Tdop)\vac}\notag\\
 = &\frac{N}{2}\left(1+(1+\mu)\Sigma_{c}\right)^{-1}\Sigma_{c}\\
 = &\frac{N}{2}\frac{\Sigma_{c}}{\left(1+\Sigma_{c}\right)}=\frac{N}{2}\left[1-\left(1+\Sigma_{c}\right)^{-1}\right]=\frac{N}{2}\left[1+G_{c}\right],
\end{align*}
which according to \eqn{eqn:rhoL-gen-func-app} gives us the density of particles on the left side
\begin{equation}
\rho_{L}=1+G_{c}.
\end{equation}
Similarly, the density of holes on the right-side, i.e. $\langle \hd_i\h_i\rangle$, can be calculated by using the source-term $\exp({\mu}\hd_i\h_i)$ in place of $\exp({\mu}\cd_ic_i)$ in \eqn{eqn:rhoL-gen-func-app} to get $\langle \hd_i\h_i\rangle=1+G_h$, from which the density of right-side fermions (particles) can be determined to be
\begin{align}
    \rho_R=1-\langle \hd_i\h_i\rangle=-G_h.
\end{align}

\section{Thermal field theory for the partitioned-SYK model}\label{sec:thermal-field-theory-pSYK}
We now discuss the the thermal field theory for the partitioned-SYK model. This will allow us to compute the exact properties for the ground-state when the temperature $T$ is extrapolated to \emph{zero}. We reiterate the Hamiltonian for the partitioned-SYK model for ease of access
\begin{align}
    \Hres=&g \sum_{
    \mathclap{\substack{i_{1}<...<i_{(q/2)} \in L,\\j_{1}<...<j_{(q/2)} \in R}}} J_{i_1\cdots i_{(q/2)};j_1\cdots j_{(q/2)}}\cd_{i_{1}}...\cd_{i_{(q/2)}} c_{j_{1}}...c_{j_{(q/2)}}\notag\\
    &+h.c.~,
\end{align}
where $g={(q/2)!}/{\sqrt{q/2}(\sqrt{N_LN_R})^{\frac{q-1}{2}}}$.
The partition function $Z=\Tr[\exp(-\beta \Hres)]$, where $\beta=T^{-1}$
, can be written as a path-integral
\begin{align}
Z=&\int\mathcal{D}[\cb,c] \exp{\left(-S[\cb,c]\right)},\notag\\
S[\cb,c]=&\int_0^\beta\dtau{\Bigg[}\sum_{i}\cb_{i}(\tau)\partial_\tau c_i(\tau)\notag\\
&+g \sum_{
    \mathclap{\substack{i_{1}<...<i_{(q/2)} \in L,\\j_{1}<...<j_{(q/2)} \in R}}}
    \left(\Jij\cb_{i_{1}}...\cb_{i_{(q/2)}} c_{j_{1}}...c_{j_{(q/2)}}\right.\notag\\
& \qquad\qquad\qquad\left.+h.c\right){\Bigg]}.
\end{align}
using the fermionic coherent states $|c_{i=1\cdots N}\rangle$ \cite{Haldar2018PRB} described by anti-periodic Grassmann fields $\cb_i(\tau)$, $c_i(\tau)$ living on the imaginary-time interval $\tau\in[0,\beta]$. In the above equation, we have defined the action $S[\cb,c]$  whose saddle-point would give us access to the large-$N$ limit. Since, we are interested in the disorder averaged free-energy 
\begin{align}
F=-T\overline{\ln Z},
\end{align}
we use the replica trick, $\overline{\ln(Z)}=\lim_{\rR\to0}(\overline{Z^\rR}-1)/\rR$, yet again, to perform the averaging over $\Jij$. The replica-partition function $\overline{Z^\rR}$ is found to be
\begin{align}
    \overline{Z^\rR}=&\int\mathcal{D}[\cb,c] \exp{\left(-S_\rR[\cb,c]\right)},\notag\\
    S_\rR=&
    \int_0^\beta\dtau_{1,2}\sum_{r_1,r_2=1}^\rR\left[\sum_{i=1}^N\cb_{ir_1}(\tau_1)\delta_{r_1,r_2}\delta(\tau_1-\tau_2)\partial_{\tau_1} c_{ir_1}(\tau_2)\right.\notag\\
    &-\frac{2J^2}{q(\sqrt{N_LN_R})^{q-1}}
\left(\sum_{i\in L}\cb_{i,r_1}(\tau_1)c_{i,r_2}(\tau_2)\right)^{q/2}\notag\\
&\left.
\qquad\qquad\qquad\qquad\left(\sum_{j\in R}c_{j,r_1}(\tau_1)\cb_{j,r_2}(\tau_2)\right)^{q/2}
\right],
\end{align}
where $S_\rR$ denotes a new action over replicas and the Grassmann fields $\cb_{i,r_1}(\tau)$, $c_{i,r_2}(\tau)$ have picked up the additional replica indices $r_1,r_2$. Introducing the large-$N$ Green's functions 
\begin{align}
    G_{L}^{(r_1,r_2)}(\tau_1,\tau_2)=&-\frac{1}{N_L}\sum_{i\in L}\langle c_{ir_1}(\tau_1)\cb_{ir_2}(\tau_2)\rangle\notag\\
    G_{R}^{(r_1,r_2)}(\tau_1,\tau_2)=&-\frac{1}{N_R}\sum_{j\in R}\langle c_{ir_1}(\tau_1)\cb_{ir_2}(\tau_2)\rangle,
\end{align}
for the left-side, right-side fermions along with their associated self-energies $\Sigma_L^{(r_1,r_2)}(\tau_1,\tau_2)$, $\Sigma_R^{(r_1,r_2)}(\tau_1,\tau_2)$, and subsequently integrating out the fermionic-fields $\cb_i$, $c_i$, etc., we arrive at the following action
\begin{widetext}
\begin{align}
    S_\rR[G,\Sigma]=&
    -N_L\ln\det\left[\delta_{r_1,r_2}\delta(\tau_1-\tau_2)\partial_{\tau_1}+\Sigma_{L}\right]
    -N_R\ln\det\left[\delta_{r_1,r_2}\delta(\tau_1-\tau_2)\partial_{\tau_1}+\Sigma_{R}\right]&\notag\\
    &-\frac{2J^2\sqrt{N_LN_R}(-1)^{q/2}}{q}
    \int_0^\beta\dtau_{1,2}\sum_{r_1,r_2=1}^\rR
 G_{L}^{(r_1,r_2)}(\tau_1,\tau_2)^{q/2}
 G_{R}^{(r_2,r_1)}(\tau_2,\tau_1)^{q/2}\notag\\
 &-\int_0^\beta\dtau_{1,2}\sum_{r_1,r_2=1}^\rR
 \left[
 N_L\Sigma_{L}^{(r_1,r_2)}(\tau_1,\tau_2)G_{L}^{(r_2,r_1)}(\tau_2,\tau_1)
 +N_R\Sigma_{R}^{(r_1,r_2)}(\tau_1,\tau_2)G_{R}^{(r_2,r_1)}(\tau_2,\tau_1)
 \right],
\end{align}
\end{widetext}
such that $\overline{Z^\rR}=\int \mathcal{D}[G,\Sigma]\exp\left(-S_\rR[G,\Sigma]\right)$. Assuming time-translational invariance and a replica-diagonal ansatz for the saddle-point, i.e. $G_{L,R}^{(r_1,r_2)}(\tau_1,\tau_2)\propto\delta_{r_1,r_2}G_{L,R}(\tau_1-\tau_2)$ and same for $\Sigma_{L,R}^{(r_1,r_2)}(\tau_1,\tau_2)$, we obtain a simplified form for the replica-action
\begin{align}\label{eqn:thermal-final-action-app}
    S_\rR=&\rR N\mathcal{S}\notag\\
    =&N(1+p)^{-1}\Big[  -\ln\det\left[\delta(\tau_1-\tau_2)\partial_{\tau_1}+\Sigma_{L}\right]\notag\\
    &-p\ln\det\left[\delta(\tau_1-\tau_2)\partial_{\tau_1}+\Sigma_{R}\right]&\notag\\
    &-\frac{2J^2\sqrt{p}(-1)^{q/2}}{q}
    \beta\int_0^\beta\dtau
 G_{L}(\tau)^{q/2}
 G_{R}(-\tau)^{q/2}\notag\\
 &-\beta\int_0^\beta\dtau\left(\Sigma_{L}(\tau)G_L(-\tau)+p\Sigma_{R}(\tau)G_R(-\tau)\right)\Big],
\end{align}
where we have used the site-ratio $p=N_R/N_L$ and defined the action-per-replica $\mathcal{S}$. The saddle-point conditions are found to be
\begin{align}\label{eqn:thermal-SP-app}
    G_{L,R}=&-\left[\delta(\tau_1-\tau_2)\partial_{\tau_1}+\Sigma_{L,R}\right]^{-1}\notag\\
    \Sigma_{L}(\tau)=&(-1)^{q/2+1}J^2\sqrt{p} G_{R}(\tau)^{q/2}
 G_{L}(-\tau)^{q/2-1}\notag\\
    \Sigma_{R}(\tau)=&(-1)^{q/2+1}\frac{J^2}{\sqrt{p}} G_{L}(\tau)^{q/2}
 G_{R}(-\tau)^{q/2-1},
\end{align}
by minimizing $\mathcal{S}$ w.r.t. $G$, $\Sigma$. The above equations were solved iteratively\cite{Haldar2018PRB}, for a given value of $T$ and $p$,  after discretizing the imaginary-time interval $[0,\beta]$. The free-energy can then be calculated by plugging the solutions of \eqn{eqn:thermal-SP-app} in \eqn{eqn:thermal-final-action-app} and using
\begin{align}
    F=-T\ln(Z)=-T\lim_{\rR\to 0}\frac{\exp^{-S[G,\Sigma]}-1}{\rR}=TN\mathcal{S}.
\end{align}
The ground-state energy density $\Etherm$ is calculated using the thermodynamic relation
\begin{align}
    \Etherm=(F/N)+T\mathbf{s},
\end{align}
where $\mathbf{s}=-(1/N)\partial_T F$ is the entropy-density obtained from $F$ via numerical differentiation. The density for the left, right side fermions are obtained using 
\begin{align}
    \rho_{L,R}=G_{L,R}(\tau=0^-),
\end{align}
which follows from the usual definition of the two-point Green's functions. We access the energy-density and particle-density for the ground-state by numerically extrapolating the values for small but finite $T$ to $T\to 0$.

\section{The non-interacting ($q=2$) partitioned-SYK model}\label{sec:one-body-p-SYK}
In this section, we study the non-interacting limit of the partitioned-SYK model on a system of $2N$ sites. The Hamiltonian for the model is obtained by setting $q=2$ in \eqn{eqn:clusop}, which is
\begin{align}
    \Hres(q=2)=&\Top +\Tdop,
\end{align}
where
\bea
 \Tdop=& \frac1{\sqrt{N}} \sum_{ij} J_{ij} c^\dagger_i d_j \\
\Top =& \frac1{\sqrt{N}} \sum_{ij} J_{ij}^* d^\dagger_j c_i. 
\eea
The single-particle spectrum of the above model was checked to be gapless via exact-diagonalization of the Hamiltonian. The energy for our variational ansatz was also calculated using the large-$N$ approach discussed in the main-text (see \appcite{SM:large-N-ana-WF}) and the minimized energy was found to be $\Emin=-0.3849...$. Interestingly, due to the non-interacting nature of the Hamiltonian, we can calculate this value for energy analytically. We now discuss the analytical approach. For simplicity, let us take $J_{ij}$ to be a real $N$ by $N$ symmetric matrix, with Gaussian matrix elements having variance $1$. We can then diagonalize $\Tdop$, leading to 
\bea
\Tdop = \sum_\mu \epsilon_\mu C^\dagger_\mu D_\mu
\eea
with $\epsilon_\mu \sim O(1)$ distributed according to semi-circle law. The operators $C_\mu^\dagger$, $C_\mu$ and $D_\mu^\dagger$, $D_\mu$ represent the single-particle eigen-states (orbitals) obtained after diagonalization and obey fermionic anti-commutation relations. We express the variational ansatz in the following way
\bea\label{eqn:VWF-q-1-app}
\VWF &= e^{-a \Tdop} \ket{0} \\
&= \bigotimes_\mu ( \ket{01}_{\mu} + (-a) \epsilon_\mu \ket{10}_{\mu}),
\eea
where $\bigotimes_\mu$ represents the direct product operation and $\ket{01}_{\mu}$  denotes the state where the $C_\mu$-orbital is occupied and the $D_\mu$-orbital is empty, while the reverse is true for the state $\ket{10}_{\mu}$.
The energy for the above ansatz is then obtained as
\bea
E/2 &= \frac{\VWFbra \Top+\Tdop \VWF}{\sand{\psi}{\psi}} \\
&= -\sum_\mu \epsilon_\mu \frac{a \epsilon_\mu}{1 + a^2 \epsilon_\mu^2} \\
&= -\frac1a \mathrm{Tr}\left[ \frac{a^2 \mathcal{J}^2}{1+a^2 \mathcal{J}^2} \right] \\
&= -\frac1{a} \sum_{n \geq 1} (-1)^{n+1} a^{2n} \mathrm{Tr}\left[\mathcal{J}^{2n}  \right] \\
&= +\frac1a \sum_{n \geq 1}  (-a^2)^{n} \mathrm{Tr}\left[\mathcal{J}^{2n}  \right] ,
\eea
where we have defined the matrix $\mathcal{J}=\frac{1}{\sqrt{N}}J_{ij}$. We can now take the disorder average using random matrix theory
\bea
\overline{\mathrm{Tr}\left[\mathcal{J}^{2n}  \right]}  = N C_n,
\eea
where $C_n$ are the Catalan numbers. Using which we find
\bea
\overline{E}/2N = -\frac1a (F(-a^2)-1)
\eea
where $F(x)$ is the ordinary generating function of the Catalan numbers, given by
\bea
F(x) = \sum_{n\geq 0} C_n x^n = \frac{1-\sqrt{1-4x}}{2x}.
\eea
This leads to an expression for the disorder-averaged energy
\bea
\overline{E}/2N =- \frac{1+2 a^2 - \sqrt{1+4 a^2}}{2 a^3}.
\eea
The minimum of $\overline{E}/2N$ occurs at $a=\sqrt{3}/2$, and the minimum value is $\Emin=\overline{E}/2N = -2/(3\sqrt{3}) \simeq -0.38$, which is equal to the value reported in the beginning of this section.
\begin{figure}
    \centering
\begin{tikzpicture}
\draw(-3.0,2.5)node{(a)};
\draw(0.0,0.0)node{\includegraphics[width=0.38\textwidth]{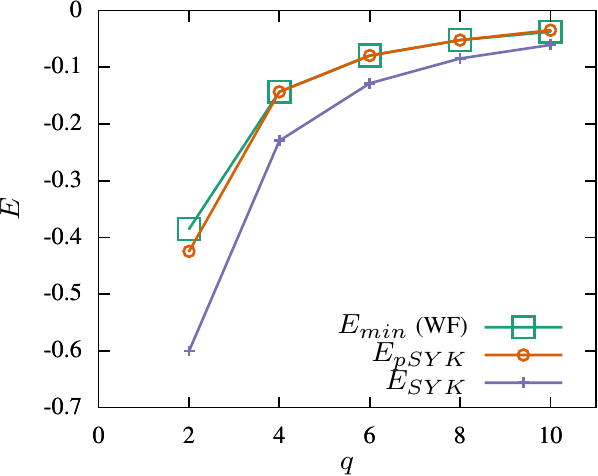}};
\draw(-3.0,-3)node{(b)};
\draw(0.0,-4.5)node{\begin{tabular}{|c|c|c|c|}
    \hline
    $q$ & $\ESYK$ & $\Emin$ & $r=\Emin/\ESYK$\\
    \hline
    2 & -0.600 & -0.385 & 0.641\\
    4 & -0.230 & -0.143 & 0.623\\
    6 & -0.128 & -0.079 & 0.618\\
    8 & -0.084 & -0.052 & 0.616\\
    10 & -0.061 & -0.037 & 0.619\\
    \hline
\end{tabular}};
\end{tikzpicture}
    \caption{(a) Comparison of the exact ground-state energy-density $\Etherm$ (circles), obtained using thermal-field theory by extrapolating $T\to 0$, with the prediction from the variational ansatz (\eqn{eqn:VWFCE}) $\Emin$ (squares). The match is excellent for $q\geq\qt$ since the partitioned-SYK model breaks PH-symmetry and develops a gap in the single-particle spectrum. However, when $q=\qo$, i.e. the non-interacting limit of the partitioned-SYK model, the single-particle spectrum is gapless and the prediction from the ansatz deviates from the exact value. The ground-state energy $\ESYK$(denoted with crosses) for the full $q$-body SYK model shown for comparison. (b) Tabular description of the data in (a) listing the numerical values for the approximation ratio $r=\Emin/\ESYK$. }
    \label{fig:Emin-EGS-q}
\end{figure}
\paragraph*{Comparison with exact ground state (GS)}
The exact ground state (GS) at half-filling is given by
\bea
\ket{GS} &=  \bigotimes_\mu ( \frac1{\sqrt{2}} \ket{01}_{\mu} - \frac1{\sqrt{2}} \mathrm{sign}(\epsilon_\mu) \ket{10}_{\mu}),
\eea
where the states $\ket{01}_{\mu}$, $\ket{10}_{\mu}$ have the same meaning as described below \eqn{eqn:VWF-q-1-app}. The ground-state energy-density is given by
\bea
\Etherm= -{\frac{1}{2N}} \sum_\mu |\epsilon_\mu|.
\eea
Since $\epsilon_\mu$ are distributed according to the semicircle law, taking the disorder average leads to 
\bea
\overline{\Etherm} &= - \frac12 \frac1{2\pi}\int_{-2}^2 d\epsilon \sqrt{4-\epsilon^2} |\epsilon| \\
&= - \frac{4}{3 \pi}\simeq -0.42.
\eea 
Comparing the energy-density of the wavefunction ($\Emin\approx-0.38$) with $\Etherm$, we find $\Emin>\Etherm$ (only slightly). More importantly, we see that, unlike $q\geq4$ case, the wavefunction does not predict the energy exactly when the ground-state is gapless, see \fig{fig:Emin-EGS-q}.

\section{Scaling of second-R\'{e}nyi entropy within the variational wavefunction}\label{sup:renyi}
In order to estimate entanglement within the variational wavefunction, we divide the system into parts A (sub-system) and B (rest) and compute the reduced density matrix $\hat{\rho}_A$ from the full density matrix
\begin{equation}
    \hat{\rho}=\VWF\VWFbra=\frac{1}{\norm}\exp{(-a\Tdop)}\vac\vacbra\exp{(-a\Top)}.
\end{equation}
We denote the fraction of sites in A as $x$. We demonstrate the computation of R\'{e}nyi-entropy for $x=0.5$, i.e. A is comprised of the left-side fermions, and report the result for arbitrary $x$ at the end. Additionally, we also set the site-ratio $p=1.0$. The reduced-density matrix $\hat{\rho}_A$ 
, when $x=0.5$, is found to be
\begin{align}
    \hat{\rho}_{A}=&\Tr_{B}[\hat{\rho}]=\int \prod_{i\in R}d\hb_{i}dh_{i} \exp\left(-\sum_i\hb_ih_i\right)\bra{-h_i}\hat{\rho}\ket{h_i}\notag\\
    =&\frac{1}{\norm}\int \prod_i d^2h_i \exp\left(-\sum_i\hb_ih_i\right)\notag\\
    &\qquad\exp{\left(-(-1)^{\q} ag J_{ij}c^{\dagger}_{i_1}...c^{\dagger}_{i_{\q}}\hb_{j_{1}}...\hb_{j_{\q}}\right)}\notag\\
    &\vac_{c}\vacbra_{c}
    \exp{\left(-agJ^*_{ij}h_{j_{\q}}...h_{j_{1}}\hat{c}_{i_{\q}}...\hat{c}_{i_{1}}\right)},
\end{align}
\Arijit{
\begin{align}
    \hat{\rho}_{A}=&\Tr_{B}[\hat{\rho}]=\int \prod_{i\in R}d\hb_{i}dh_{i} \exp\left(-\sum_{i\in B}\hb_ih_i\right)\bra{-h_i}\hat{\rho}\ket{h_i}\notag\\
    =&\frac{1}{\norm}\int \prod_i d^2h_i \exp\left(-\sum_i\hb_ih_i\right)\notag\\
    &\qquad\qquad\exp{\left(-(-1)^{\q} ag J_{ij}c^{\dagger}_{i_1}...c^{\dagger}_{i_{\q}}\hd_{j}...\hb_{j_{1}}...\hb_{j_{\q}}\right)}
    \vac_{c}\vacbra_{c}
    \exp{\left(-agJ^*_{ij}h_{j_{\q}}...h_{j_{1}}\h_{j}...\hat{c}_{i_{\q}}...\hat{c}_{i_{1}}\right)},\notag
\end{align}
}
where we have used the fermionic-coherent state $|h_{i\in R}\rangle$ for the holes and their corresponding Grassmann numbers $\hb_i$, $h_i$. The symbol $\vac_c$ denotes the vacuum for the $\co_{i\in L}$ fermions. Since, the $2$-nd R\'{e}nyi-entropy is related to the second moment of the reduced-density matrix, i.e. 
\begin{align}\label{eqn:EE-def-app}
    \EE=-\overline{\ln\Tr[\hat{\rho}_A^2]},
\end{align}
we represent $\hat{\rho}^{2}_{A}$ as an integral over Grassmann-variables

\begin{align}
    \hat{\rho}^{2}_{A} &=\frac{1}{\norm^2}
    {\bigg[}\int d^2h_{J1} \exp{\left(-\hb_{j1}h_{j1}\right)}\notag\\
    &\qquad\quad\qquad\quad\exp\left(-(-1)^{\q}gaJ_{IJ}\cd_{I}\hb_{J1}\right)\notag\\
    &\qquad\quad\qquad\quad\qquad\quad\vac_{c}\vacbra_{c}\exp{\left(-gaJ^{*}_{IJ}h_{J1}\co_{I}\right)}{\bigg]}\notag\\
    &\qquad\quad {\bigg[}\int d^2h_{J2}
    \exp{\left(-\hb_{j2}h_{j2}\right)}\notag\\
    &\qquad\quad\qquad\quad\exp\left(-(-1)^{\q}gaJ_{IJ}\cd_{I}\hb_{J2}\right)\notag\\
    &\qquad\quad\qquad\quad\qquad\qquad\vac_{c}\vacbra_{c}\exp{\left(-gaJ^{*}_{IJ}h_{J2}\co_{I}\right)}{\bigg]},
\end{align}
\Arijit{
\begin{align}
    \hat{\rho}^{2}_{A} &=\frac{1}{\norm^2}
    \left[\int d^2h_{J1} \exp{\left(-\hb_{j1}h_{j1}\right)}
    \exp\left(-(-1)^{\q}gaJ_{IJ}\cd_{I}\hd_{j}..\hb_{J1}\right)\vac_{c}\vacbra_{c}\exp{\left(-gaJ^{*}_{IJ}h_{J1}\h_{j}..\co_{I}\right)}\right]\notag\\
    &\qquad\quad \left[\int d^2h_{J2}
    \exp{\left(-\hb_{j2}h_{j2}\right)}
    \exp\left(-(-1)^{\q}gaJ_{IJ}\cd_{I}\hd_{j}..\hb_{J2}\right)\vac_{c}\vacbra_{c}\exp{\left(-gaJ^{*}_{IJ}h_{J2}\h_j..\co_{I}\right)}\right],\notag
\end{align}}
where we have introduced the shorthand notation $I=\{i_{1},...,i_{\q}\}$, $J=\{j_{1},...,j_{\q}\}$, $c_{I}=c_{i_{1}}...c_{i_{\q}}$ etc., with sum over repeated indices ($i$, $j$, $I$, $J$) implied. We can evaluate the trace of $\hat{\rho}_A^2$ by introducing the Grassmann numbers $\cb_{i1}$, $c_{i1}$, to get

\begin{align}\label{eqn:Trrho2-step-1-app}
&\Tr_{A}(\hat{\rho}_{A}^2)\notag\\
    =&\frac{1}{\norm^2}
    \int d^2c_{I1}
    \exp{\left(-\cb_{i1}c_{i1}\right)}\notag\\
    &{\bigg[}
    \int d^2h_{J1}
    \exp{\left(-\hb_{j1}h_{j1}\right)}
    \exp\left((-1)^{\tq+1}gaJ_{IJ}\cb_{I1}\hb_{J1}\right)\notag\\
    &\qquad\qquad\qquad\qquad\vacbra_{c}\exp(-gaJ^{*}_{IJ}h_{J1}\co_{I}){\bigg]}\notag\\
    &{\bigg[}\int d^2h_{J2}\exp(-\hb_{j2}h_{j2})
    \exp\left(-(-1)^{\q}agJ_{IJ}\cd_{I}\hb_{J2}\right)
    \notag\\
    &\qquad\qquad\qquad\qquad\vac_{c}
    \exp\left(-agJ^{*}_{IJ}h_{J2}c_{I1}\right){\bigg]}.
\end{align}
\newcommand{\htil}{\tilde{h}}
\newcommand{\htilb}{\bar{\tilde{h}}}
\Arijit{
\begin{align}\label{eqn:Trrho2-step-1-app}
    \Tr_{A}(\hat{\rho}_{A}^2)&=\frac{1}{\norm^2}
    \int d^2[c_{I1},\htil_{A1}]
    \exp{\left(-\cb_{i1}c_{i1}-\htilb_{j1}\htil_{j1}\right)}\notag\\
    &
    \left[
    \int d^2h_{J1}
    \exp{\left(-\hb_{j1}h_{j1}\right)}
    \exp\left((-1)^{\tq+1}gaJ_{IJ}\cb_{I1}\htilb_{j1}\hb_{J1}\right)
    \vacbra_{c}\exp(-gaJ^{*}_{IJ}h_{J1}\h_{j}\co_{I})\right]\notag\\
    &\left[\int d^2h_{J2}\exp(-\hb_{j2}h_{j2})
    \exp\left(-(-1)^{\q}agJ_{IJ}\cd_{I}\hd_{J}\hb_{J2}\right)\vac_{c}
    \exp\left(-agJ^{*}_{IJ}h_{J2}\htil_{J1}c_{I1}\right)\right].\notag
\end{align}}
The expectation value appearing inside the trace can be evaluated by introducing $\cb_{i2}$, $c_{i2}$, as shown below 
\begin{align}
    &\vacbra_{c}
    \exp\left(-agJ^{*}_{IJ}h_{J1}\co_{I}\right)
    \exp\left((-1)^{1+\q}agJ_{IJ}\cd_{I}\hb_{J2}\right)
    \vac_{c}\notag\\
    =&\int d^2c_{I2}
    \exp(-\cb_{i2}c_{i2})
    \vacbra_{c}
    \exp(-agJ^{*}_{IJ}h_{J1}\co_{I})
    |c_{i2}\rangle\notag\\
    &\qquad
    \langle c_{i2}|
    \exp\left(-(-1)^{\q}agJ_{IJ}\cd_{I}\hb_{J2}\right)
    \vac_{c}\notag\\
    =&\int d^2c_{I2}
    \exp(-\cb_{i2}c_{i2})
    \exp\left(-gaJ^{*}_{IJ}h_{J1}c_{I2}\right)\notag\\
    &\qquad
    \exp\left(-(-1)^{\q}agJ_{IJ}\cb_{I2}\hb_{J2}\right)
\end{align}
Substituting the above expression into \eqn{eqn:Trrho2-step-1-app}, we get an expression for $\Tr_{A}(\hat{\rho}_{A}^2)$ involving \emph{only} Grassmann-variables, i.e.
\begin{equation}
    \Tr_{A}[\hat{\rho}^{2}_{A}]=\frac{1}{\norm^2}
    \int d^2c_{I1,I2}d^2h_{J1,J2} \exp(-\mathcal{S}),
\end{equation}
where the action
\begin{align}
    \mathcal{S}=&
    \sum_{\alpha=1}^{2}(-\cb_{i\alpha}c_{i\alpha}-\hb_{j\alpha}h_{j\alpha})\notag\\
    &+ga\sum_{I,J}
    (J_{IJ}\cb_{I1}\hb_{J1}+J^{*}_{IJ}h_{J1}c_{I2}\notag\\
    &
    +(-1)^{\q}J_{IJ}\cb_{I2}h_{J2}
    +J^{*}_{IJ}h_{J2}c_{I1}).
\end{align}
A useful point to note here is that the Grassmann-variables are now indexed by an additional number $1$ or $2$, since we are dealing with the square of the density-matrix $\hat{\rho}_A$. The disorder averaging of $\ln\mathrm{Tr}[\hat{\rho}_A^2]$ (see \eqn{eqn:EE-def-app}) over $\Jij$ can be implemented in the same way, using the replica-trick, as was done for $\ln\norm$ (see \appcite{SM:large-N-ana-WF}). Subsequently, the large-$N$ limit of the resulting replica-action, like in the $\ln\norm$ case, can also be obtained by introducing the static Green's functions ($G_{c,h}$, see \eqn{eqn:s-GF-app}) and self-energies ($\Sigma_{c,h}$), except this time they are $2\times 2$ matrices. Therefore, we have
\begin{align}
    G_c=&N_L^{-1}\left[
    \begin{array}{cc}
        \sum\limits_{i\in L}\langle\cb_{i1}c_{i1}\rangle &\sum\limits_{i\in L}\langle\cb_{i1}c_{i2}\rangle  \\
        \sum\limits_{i\in L}\langle\cb_{i2}c_{i1}\rangle & \sum\limits_{i\in L}\langle\cb_{i2}c_{i2}\rangle
    \end{array}
    \right],\notag\\
        G_h=&N_R^{-1}\left[
    \begin{array}{cc}
        \sum\limits_{i\in R}\langle\hb_{i1}h_{i1}\rangle &\sum\limits_{i\in R}\langle\hb_{i1}h_{i2}\rangle  \\
        \sum\limits_{i\in R}\langle\hb_{i2}h_{i1}\rangle & \sum\limits_{i\in R}\langle\hb_{i2}h_{i2}\rangle
    \end{array}
    \right],
\end{align}
\Arijit{
\begin{align}
    G_A=N_L^{-1}\left[
    \begin{array}{cc}
        \sum\limits_{i\in L}\langle\cb_{i1}c_{i1}\rangle &\sum\limits_{i\in L}\langle\cb_{i1}c_{i2}\rangle  \\
        \sum\limits_{i\in L}\langle\cb_{i2}c_{i1}\rangle & \sum\limits_{i\in L}\langle\cb_{i2}c_{i2}\rangle
    \end{array}
    \right],
    &G_B=((1-x)N_R)^{-1}\left[
    \begin{array}{cc}
        \sum\limits_{i\in R}\langle\hb_{i1}h_{i1}\rangle &\sum\limits_{i\in R}\langle\hb_{i1}h_{i2}\rangle  \\
        \sum\limits_{i\in R}\langle\hb_{i2}h_{i1}\rangle & \sum\limits_{i\in R}\langle\hb_{i2}h_{i2}\rangle
    \end{array}
    \right],
    &\tilde{G}_A=(xN_R)^{-1}\left[
    \begin{array}{cc}
        \sum\limits_{i\in A}\langle\hb_{i1}h_{i1}\rangle &\sum\limits_{i\in A}\langle\hb_{i1}h_{i2}\rangle  \\
        \sum\limits_{i\in R}\langle\hb_{i2}h_{i1}\rangle & \sum\limits_{i\in A}\langle\hb_{i2}h_{i2}\rangle
    \end{array}
    \right],\notag
\end{align}
}
and similar definitions for the self-energies.
At the large-$N$ saddle-point, the $2$-nd R\'{e}nyi entropy is found to be
\begin{equation}
    S^{(2)}=F_1+2\overline{\ln(\norm)},
\end{equation}
where
\begin{align}\label{eqn:F1-app}
    F_1 =&-\frac{N}{2}
    {\bigg[}\ln[\det(\mathbf{1}+\Sigma_{c})]+\ln[\det(\mathbf{1}+\Sigma_{h})]\notag\\
    &+\Tr[\Sigma_c G_c]+\Tr[\Sigma_h G_h]{\bigg]}\notag\\&
    -\frac{a^2N}{\tq}\left[G_c(2,1)^{\q}G_h(1,1)^{\q}\right.\notag\\
    &\qquad\qquad+(-1)^{\q}G_c(2,2)^{\q}G_h(1,2)^{\q}\notag\\
    &\qquad\qquad+G_c(1,1)^{\q}G_h(2,1)^{\q}\notag\\
    &\qquad\qquad\left.+(-1)^{\q}G_c(1,2)^{\q}G_h(2,2)^{\q}\right],
\end{align}
where $\mathbf{1}$ represent the $2\times 2$ identity matrix and $\ln(\norm)$ is obtained from \eqn{eqn:lnNSP-app}. Also, we have set $J=1$. The $2\times 2$ matrices $G$, $\Sigma$,  appearing above, are found by solving the saddle-point conditions
\begin{align}\label{eqn:SP-F1-app}
(\mathbf{1}+\Sigma_{c,h})=&-G_{c,h}^{-1}\notag\\
    \Sigma_{c}(1,1)	=&-a^{2}G_{h}(2,1)^{\q}G_{c}(1,1)^{\q-1}\notag\\
\Sigma_{c}(1,2)	=&-a^{2}G_{h}(1,1)^{\q}G_{c}(2,1)^{\q-1}\notag\\
\Sigma_{c}(2,1)	=&-a^{2}(-1)^{\q}G_{h}(2,2)^{\q}G_{c}(1,2)^{\q-1}\notag\\
\Sigma_{c}(2,2)	=&-a^{2}(-1)^{\q}G_{h}(1,2)^{\q}G_{c}(2,2)^{\q-1},
\end{align}
with the equations for the components of $\Sigma_h$ obtained by interchanging $c\xleftrightarrow{} h$ in the subscript. Similarly, the result for arbitrary sub-system sizes $x$ can be found to be
\begin{align}
    \EE(x)=2\overline{\ln(\norm)}+
    \begin{cases}
    F(x)&0\le x\le 0.5\\
    F(1-x)&0.5\le x\le 1
    \end{cases}\ ,
\end{align}
where \Arijit{$\eta\to G_A,\ \xi\to \tilde{G}_B,\ \gamma\to G_B$}
\newcommand{\I}{\mathbf{1}}
\newcommand{\GtB}{\tilde{G}_B}
\newcommand{\StB}{\tilde{\Sigma}_B}
\begin{align}\label{eqn:Fx-app}
    & F(x)\notag\\
    =&-\frac{N}{2}
    {\bigg[}2x\ln[\det(\I+\Sigma_A)]\notag\\
    &+(1-2x)\ln[\det(\I+\StB)]
    +\ln[\det(\I+\Sigma_B)]\notag\\
    &\left.
    +2x\Tr[\Sigma_A G_A]
    +(1-2x)\Tr[\StB\GtB]
    +\Tr[\Sigma_B G_B]
    \right]\notag\\
   &
    -\frac{a^2N}{\tq}
    \left[
    \left(2x G_A(2,1)+(1-2x)\GtB(1,1)\right)^{\qf}G_B(1,1)^{\qf}\right.\notag\\
    &\left.
    +(-1)^{\qf}\left(2xG_A(2,2)-(1-2x)\GtB(1,2)\right)^{\qf}G_B(1,2)^{\qf}
    \right.\notag\\
    &
    +\left(2xG_A(1,1)+(1-2x)\GtB(2,1)\right)^{\qf}G_B(2,1)^{\qf}\notag\\
    &
    +\left.(-1)^{\qf}\left(2xG_A(1,2)-(1-2x)\GtB(2,2)\right)^{\qf}G_B(2,2)^{\qf}
    \right],
\end{align}
\Arijit{
\begin{align}\label{eqn:Fx-app}
    F_h (x)=&-\frac{N}{2}
    \left[\ln[\det(\I+\Sigma_A)]
    +x\ln[\det(\I+\tilde{\Sigma}_A)]
    +(1-x)\ln[\det(\I+\Sigma_B)]\right.\notag\\
    &\qquad\qquad\qquad\qquad\left.
    +\Tr[\Sigma_A G_A]
    +x\Tr[\tilde{\Sigma}_{A}\tilde{G}_{A}]
    +(1-x)\Tr[\Sigma_B G_B]
    \right]\notag\\
   &
    -\frac{a^2N}{\tq}
    \left[
    G_A(2,1)^{\qf}(x\tilde{G}_A(2,1)+(1-x)G_B(1,1))^{\qf}
    +(-1)^{\qf}G_A(2,2)(-x\tilde{G}_A(2,2)+(1-x)G_B(1,2))^{\qf}
    \right.\notag\\
    &
    +\left.G_A(1,1)^{\qf}(x\tilde{G}_A(1,1)+(1-x)G_B(2,1))^{\qf}
    +(-1)^{\qf}G_A(1,2)^{\qf}(-x\tilde{G}_A(1,2)+(1-x)G_B(2,2))^{\qf}
    \right],\notag
\end{align}
}
\Arijit{
\begin{enumerate}  
    \item $F_h(x)$ is obtained by integrating a portion $(1-x)$ of the right-side holes. Like $\tilde{G}_B$, which contains left-side electrons that belong to B, $\tilde{G}_A$ describes right-side holes that belong to the sub-system A.
    \item Notice that $F(x)=F_h(1-x)$, provided we interchange $G_A$ with $G_B$ and $\tilde{G}_A$ with $\tilde{G}_B$.
    \item Assuming the saddle-point equations have a unique solution. The solution $G_A$, $\tilde{G}_B$, $G_B$ for minimizing $F(x)$, should solve the saddle-point equations for minimizing $F_h(1-x)$, s.t. $G_A|_{F_h}=G_B$, $G_B|_{F_h}=G_A$, $\tilde{G}_A|_{F_h}=\tilde{G}_B$.
\end{enumerate}
}
 and the saddle-point conditions are given by
 \newcommand{\qtil}{\tilde{q}}
\begin{align}
    G_{A,B}=&-(\I+\Sigma_{A,B})^{-1}\notag\\
    \GtB=&-(\I+\StB)^{-1}\notag
\end{align}
\begin{align}
    {\Sigma^{11}_A}=&-a^{2}G_B(2,1)^{\qtil}{\big(}2xG_A(1,1)\notag\\&\qquad\qquad\qquad\qquad+(1-2x)\GtB(2,1){\big)}^{\qtil-1}\notag\\
    {\Sigma^{12}_A}=&-a^{2}G_B(1,1)^{\qtil}{\big(}2xG_A(2,1)\notag\\&\qquad\qquad\qquad\qquad+(1-2x)\GtB(1,1){\big)}^{\qtil-1}\notag\\
    {\Sigma^{21}_A}=&-a^{2}(-G_B(2,2))^{\qtil}{\big(}2xG_A(1,2)\notag\\&\qquad\qquad\qquad\qquad-(1-2x)\GtB(2,2){\big)}^{\qtil-1}\notag\\
    {\Sigma^{22}_A}=&-a^{2}(-G_B(1,2))^{\qtil}{\big(}2xG_A(2,2)\notag\\&\qquad\qquad\qquad\qquad-(1-2x)\GtB(1,2){\big)}^{\qtil-1}\notag
\end{align}
\begin{align}
    {\StB}^{11}=&-a^{2}G_B(1,1)^{\qtil}{\big(}2xG_A(2,1)\notag\\&\qquad\qquad\qquad\qquad+(1-2x)\GtB(1,1){\big)}^{\qtil-1}\notag\\
    {\StB}^{12}=&-a^{2}G_B(2,1)^{\qtil}{\big(}2xG_A(1,1)\notag\\&\qquad\qquad\qquad\qquad+(1-2x)\GtB(2,1){\big)}^{\qtil-1}\notag\\
    {\StB}^{21}=&-a^{2}(-G_B(1,2))^{\qtil}{\big(}2xG_A(2,2)\notag\\&\qquad\qquad\qquad\qquad-(1-2x)\GtB(1,2){\big)}^{\qtil-1}\notag\\
    {\StB}^{21}=&-a^{2}(-G_B(2,2))^{\qtil}{\big(}2xG_A(1,2)\notag\\&\qquad\qquad\qquad\qquad-(1-2x)\GtB(2,2){\big)}^{\qtil-1}\notag
\end{align}
\begin{align}\label{eqn:SP-Fx-app}
    {\Sigma^{11}_B}=&-a^{2}G_B(1,1)^{\qtil-1}{\big(}2xG_A(2,1)\notag\\&\qquad\qquad\qquad\qquad+(1-2x)\GtB(1,1){\big)}^{\qtil}\notag\\
    {\Sigma^{12}_B}=&-a^{2}G_B(2,1)^{\qtil-1}{\big(}2xG_A(1,1)\notag\\&\qquad\qquad\qquad\qquad+(1-2x)\GtB(2,1){\big)}^{\qtil}\notag\\
    {\Sigma^{21}_B}=&-a^{2}(-G_B(1,2))^{\qtil-1}{\big(}2xG_A(2,2)\notag\\&\qquad\qquad\qquad\qquad-(1-2x)\GtB(1,2){\big)}^{\qtil}\notag\\
    {\Sigma^{22}_B}=&-a^{2}(-G_B(2,2))^{\qtil-1}{\big(}2xG_A(1,2)\notag\\&\qquad\qquad\qquad\qquad-(1-2x)\GtB(2,2){\big)}^{\qtil},
\end{align}
where we have defined $\qtil=\q$. Here $G_A$ ($\GtB$) represents the Green's function for the left-side fermions in A (B) and $G_B$ the Green's function for the right-side fermions in B. The same convention applies for the self-energies as well. When we substitute $x=0.5$ into \eqns{eqn:Fx-app}{eqn:SP-Fx-app}, $G,\Sigma_{A,B}\to G,\Sigma_{c,h}$ while the terms involving $\GtB$ drop out, and we recover \eqns{eqn:F1-app}{eqn:SP-F1-app} respectively.

\fi
\bibliography{bibliography}

\begin{thebibliography}{71}%
\makeatletter
\providecommand \@ifxundefined [1]{%
 \@ifx{#1\undefined}
}%
\providecommand \@ifnum [1]{%
 \ifnum #1\expandafter \@firstoftwo
 \else \expandafter \@secondoftwo
 \fi
}%
\providecommand \@ifx [1]{%
 \ifx #1\expandafter \@firstoftwo
 \else \expandafter \@secondoftwo
 \fi
}%
\providecommand \natexlab [1]{#1}%
\providecommand \enquote  [1]{``#1''}%
\providecommand \bibnamefont  [1]{#1}%
\providecommand \bibfnamefont [1]{#1}%
\providecommand \citenamefont [1]{#1}%
\providecommand \href@noop [0]{\@secondoftwo}%
\providecommand \href [0]{\begingroup \@sanitize@url \@href}%
\providecommand \@href[1]{\@@startlink{#1}\@@href}%
\providecommand \@@href[1]{\endgroup#1\@@endlink}%
\providecommand \@sanitize@url [0]{\catcode `\\12\catcode `\$12\catcode
  `\&12\catcode `\#12\catcode `\^12\catcode `\_12\catcode `\%12\relax}%
\providecommand \@@startlink[1]{}%
\providecommand \@@endlink[0]{}%
\providecommand \url  [0]{\begingroup\@sanitize@url \@url }%
\providecommand \@url [1]{\endgroup\@href {#1}{\urlprefix }}%
\providecommand \urlprefix  [0]{URL }%
\providecommand \Eprint [0]{\href }%
\providecommand \doibase [0]{http://dx.doi.org/}%
\providecommand \selectlanguage [0]{\@gobble}%
\providecommand \bibinfo  [0]{\@secondoftwo}%
\providecommand \bibfield  [0]{\@secondoftwo}%
\providecommand \translation [1]{[#1]}%
\providecommand \BibitemOpen [0]{}%
\providecommand \bibitemStop [0]{}%
\providecommand \bibitemNoStop [0]{.\EOS\space}%
\providecommand \EOS [0]{\spacefactor3000\relax}%
\providecommand \BibitemShut  [1]{\csname bibitem#1\endcsname}%
\let\auto@bib@innerbib\@empty
%</preamble>
\bibitem [{\citenamefont {Laughlin}(1983)}]{Laughlin}%
  \BibitemOpen
  \bibfield  {author} {\bibinfo {author} {\bibfnamefont {R.~B.}\ \bibnamefont
  {Laughlin}},\ }\bibfield  {title} {\enquote {\bibinfo {title} {Anomalous
  quantum hall effect: An incompressible quantum fluid with fractionally
  charged excitations},}\ }\href {\doibase 10.1103/PhysRevLett.50.1395}
  {\bibfield  {journal} {\bibinfo  {journal} {Phys. Rev. Lett.}\ }\textbf
  {\bibinfo {volume} {50}},\ \bibinfo {pages} {1395--1398} (\bibinfo {year}
  {1983})}\BibitemShut {NoStop}%
\bibitem [{\citenamefont {Bardeen}\ \emph {et~al.}(1957)\citenamefont
  {Bardeen}, \citenamefont {Cooper},\ and\ \citenamefont {Schrieffer}}]{BCS}%
  \BibitemOpen
  \bibfield  {author} {\bibinfo {author} {\bibfnamefont {J.}~\bibnamefont
  {Bardeen}}, \bibinfo {author} {\bibfnamefont {L.~N.}\ \bibnamefont {Cooper}},
  \ and\ \bibinfo {author} {\bibfnamefont {J.~R.}\ \bibnamefont {Schrieffer}},\
  }\bibfield  {title} {\enquote {\bibinfo {title} {Theory of
  superconductivity},}\ }\href {\doibase 10.1103/PhysRev.108.1175} {\bibfield
  {journal} {\bibinfo  {journal} {Phys. Rev.}\ }\textbf {\bibinfo {volume}
  {108}},\ \bibinfo {pages} {1175--1204} (\bibinfo {year} {1957})}\BibitemShut
  {NoStop}%
\bibitem [{\citenamefont {Gutzwiller}(1963)}]{Gutzwiller}%
  \BibitemOpen
  \bibfield  {author} {\bibinfo {author} {\bibfnamefont {Martin~C.}\
  \bibnamefont {Gutzwiller}},\ }\bibfield  {title} {\enquote {\bibinfo {title}
  {Effect of correlation on the ferromagnetism of transition metals},}\ }\href
  {\doibase 10.1103/PhysRevLett.10.159} {\bibfield  {journal} {\bibinfo
  {journal} {Phys. Rev. Lett.}\ }\textbf {\bibinfo {volume} {10}},\ \bibinfo
  {pages} {159--162} (\bibinfo {year} {1963})}\BibitemShut {NoStop}%
\bibitem [{\citenamefont {Laflorencie}(2016)}]{Entanglement}%
  \BibitemOpen
  \bibfield  {author} {\bibinfo {author} {\bibfnamefont {Nicolas}\ \bibnamefont
  {Laflorencie}},\ }\bibfield  {title} {\enquote {\bibinfo {title} {Quantum
  entanglement in condensed matter systems},}\ }\href {\doibase
  https://doi.org/10.1016/j.physrep.2016.06.008} {\bibfield  {journal}
  {\bibinfo  {journal} {Physics Reports}\ }\textbf {\bibinfo {volume} {646}},\
  \bibinfo {pages} {1 -- 59} (\bibinfo {year} {2016})}\BibitemShut {NoStop}%
\bibitem [{\citenamefont {Feynman}(1982)}]{Feynman}%
  \BibitemOpen
  \bibfield  {author} {\bibinfo {author} {\bibfnamefont {Richard~P.}\
  \bibnamefont {Feynman}},\ }\bibfield  {title} {\enquote {\bibinfo {title}
  {Simulating physics with computers},}\ }\href {\doibase 10.1007/BF02650179}
  {\bibfield  {journal} {\bibinfo  {journal} {International Journal of
  Theoretical Physics}\ }\textbf {\bibinfo {volume} {21}},\ \bibinfo {pages}
  {467--488} (\bibinfo {year} {1982})}\BibitemShut {NoStop}%
\bibitem [{\citenamefont {Lloyd}(1996)}]{Lloyd1073}%
  \BibitemOpen
  \bibfield  {author} {\bibinfo {author} {\bibfnamefont {Seth}\ \bibnamefont
  {Lloyd}},\ }\bibfield  {title} {\enquote {\bibinfo {title} {Universal quantum
  simulators},}\ }\href {\doibase 10.1126/science.273.5278.1073} {\bibfield
  {journal} {\bibinfo  {journal} {Science}\ }\textbf {\bibinfo {volume}
  {273}},\ \bibinfo {pages} {1073--1078} (\bibinfo {year} {1996})},\ \Eprint
  {http://arxiv.org/abs/https://science.sciencemag.org/content/273/5278/1073.full.pdf}
  {https://science.sciencemag.org/content/273/5278/1073.full.pdf} \BibitemShut
  {NoStop}%
\bibitem [{\citenamefont {McClean}\ \emph {et~al.}(2016)\citenamefont
  {McClean}, \citenamefont {Romero}, \citenamefont {Babbush},\ and\
  \citenamefont {Aspuru-Guzik}}]{McClean_2016}%
  \BibitemOpen
  \bibfield  {author} {\bibinfo {author} {\bibfnamefont {Jarrod~R}\
  \bibnamefont {McClean}}, \bibinfo {author} {\bibfnamefont {Jonathan}\
  \bibnamefont {Romero}}, \bibinfo {author} {\bibfnamefont {Ryan}\ \bibnamefont
  {Babbush}}, \ and\ \bibinfo {author} {\bibfnamefont {Al{\'{a}}n}\
  \bibnamefont {Aspuru-Guzik}},\ }\bibfield  {title} {\enquote {\bibinfo
  {title} {The theory of variational hybrid quantum-classical algorithms},}\
  }\href {\doibase 10.1088/1367-2630/18/2/023023} {\bibfield  {journal}
  {\bibinfo  {journal} {New Journal of Physics}\ }\textbf {\bibinfo {volume}
  {18}},\ \bibinfo {pages} {023023} (\bibinfo {year} {2016})}\BibitemShut
  {NoStop}%
\bibitem [{\citenamefont {Kandala}\ \emph {et~al.}(2017)\citenamefont
  {Kandala}, \citenamefont {Mezzacapo}, \citenamefont {Temme}, \citenamefont
  {Takita}, \citenamefont {Brink}, \citenamefont {Chow},\ and\ \citenamefont
  {Gambetta}}]{IBM}%
  \BibitemOpen
  \bibfield  {author} {\bibinfo {author} {\bibfnamefont {Abhinav}\ \bibnamefont
  {Kandala}}, \bibinfo {author} {\bibfnamefont {Antonio}\ \bibnamefont
  {Mezzacapo}}, \bibinfo {author} {\bibfnamefont {Kristan}\ \bibnamefont
  {Temme}}, \bibinfo {author} {\bibfnamefont {Maika}\ \bibnamefont {Takita}},
  \bibinfo {author} {\bibfnamefont {Markus}\ \bibnamefont {Brink}}, \bibinfo
  {author} {\bibfnamefont {Jerry~M.}\ \bibnamefont {Chow}}, \ and\ \bibinfo
  {author} {\bibfnamefont {Jay~M.}\ \bibnamefont {Gambetta}},\ }\bibfield
  {title} {\enquote {\bibinfo {title} {Hardware-efficient variational quantum
  eigensolver for small molecules and quantum magnets},}\ }\href {\doibase
  10.1038/nature23879} {\bibfield  {journal} {\bibinfo  {journal} {Nature}\
  }\textbf {\bibinfo {volume} {549}},\ \bibinfo {pages} {242--246} (\bibinfo
  {year} {2017})}\BibitemShut {NoStop}%
\bibitem [{HF_(2020)}]{HF_Google}%
  \BibitemOpen
  \bibfield  {title} {\enquote {\bibinfo {title} {Hartree-fock on a
  superconducting qubit quantum computer},}\ }\href {\doibase
  10.1126/science.abb9811} {\bibfield  {journal} {\bibinfo  {journal}
  {Science}\ }\textbf {\bibinfo {volume} {369}},\ \bibinfo {pages} {1084--1089}
  (\bibinfo {year} {2020})},\ \Eprint
  {http://arxiv.org/abs/https://science.sciencemag.org/content/369/6507/1084.full.pdf}
  {https://science.sciencemag.org/content/369/6507/1084.full.pdf} \BibitemShut
  {NoStop}%
\bibitem [{\citenamefont {Gharibian}\ \emph {et~al.}(2015)\citenamefont
  {Gharibian}, \citenamefont {Huang}, \citenamefont {Landau},\ and\
  \citenamefont {Shin}}]{TCS-066}%
  \BibitemOpen
  \bibfield  {author} {\bibinfo {author} {\bibfnamefont {Sevag}\ \bibnamefont
  {Gharibian}}, \bibinfo {author} {\bibfnamefont {Yichen}\ \bibnamefont
  {Huang}}, \bibinfo {author} {\bibfnamefont {Zeph}\ \bibnamefont {Landau}}, \
  and\ \bibinfo {author} {\bibfnamefont {Seung~Woo}\ \bibnamefont {Shin}},\
  }\bibfield  {title} {\enquote {\bibinfo {title} {Quantum hamiltonian
  complexity},}\ }\href {\doibase 10.1561/0400000066} {\bibfield  {journal}
  {\bibinfo  {journal} {Foundations and Trends{\textregistered} in Theoretical
  Computer Science}\ }\textbf {\bibinfo {volume} {10}},\ \bibinfo {pages}
  {159--282} (\bibinfo {year} {2015})}\BibitemShut {NoStop}%
\bibitem [{\citenamefont {Kitaev}\ \emph {et~al.}(2002)\citenamefont {Kitaev},
  \citenamefont {Shen}, \citenamefont {Vyalyi},\ and\ \citenamefont
  {Vyalyi}}]{kitaev2002classical}%
  \BibitemOpen
  \bibfield  {author} {\bibinfo {author} {\bibfnamefont {Alexei~Yu}\
  \bibnamefont {Kitaev}}, \bibinfo {author} {\bibfnamefont {Alexander}\
  \bibnamefont {Shen}}, \bibinfo {author} {\bibfnamefont {Mikhail~N}\
  \bibnamefont {Vyalyi}}, \ and\ \bibinfo {author} {\bibfnamefont {Mikhail~N}\
  \bibnamefont {Vyalyi}},\ }\href@noop {} {\emph {\bibinfo {title} {Classical
  and quantum computation}}},\ \bibinfo {number} {47}\ (\bibinfo  {publisher}
  {American Mathematical Soc.},\ \bibinfo {year} {2002})\BibitemShut {NoStop}%
\bibitem [{\citenamefont {Kempe}\ \emph {et~al.}(2005)\citenamefont {Kempe},
  \citenamefont {Kitaev},\ and\ \citenamefont
  {Regev}}]{10.1007/978-3-540-30538-5_31}%
  \BibitemOpen
  \bibfield  {author} {\bibinfo {author} {\bibfnamefont {Julia}\ \bibnamefont
  {Kempe}}, \bibinfo {author} {\bibfnamefont {Alexei}\ \bibnamefont {Kitaev}},
  \ and\ \bibinfo {author} {\bibfnamefont {Oded}\ \bibnamefont {Regev}},\
  }\bibfield  {title} {\enquote {\bibinfo {title} {The complexity of the local
  hamiltonian problem},}\ }in\ \href@noop {} {\emph {\bibinfo {booktitle}
  {FSTTCS 2004: Foundations of Software Technology and Theoretical Computer
  Science}}},\ \bibinfo {editor} {edited by\ \bibinfo {editor} {\bibfnamefont
  {Kamal}\ \bibnamefont {Lodaya}}\ and\ \bibinfo {editor} {\bibfnamefont
  {Meena}\ \bibnamefont {Mahajan}}}\ (\bibinfo  {publisher} {Springer Berlin
  Heidelberg},\ \bibinfo {address} {Berlin, Heidelberg},\ \bibinfo {year}
  {2005})\ pp.\ \bibinfo {pages} {372--383}\BibitemShut {NoStop}%
\bibitem [{\citenamefont {Watrous}(2008)}]{watrous2008quantum}%
  \BibitemOpen
  \bibfield  {author} {\bibinfo {author} {\bibfnamefont {John}\ \bibnamefont
  {Watrous}},\ }\bibfield  {title} {\enquote {\bibinfo {title} {Quantum
  computational complexity},}\ }\href@noop {} {\bibfield  {journal} {\bibinfo
  {journal} {arXiv preprint arXiv:0804.3401}\ } (\bibinfo {year}
  {2008})}\BibitemShut {NoStop}%
\bibitem [{\citenamefont {Gharibian}\ and\ \citenamefont
  {Kempe}(2012)}]{doi:10.1137/110842272}%
  \BibitemOpen
  \bibfield  {author} {\bibinfo {author} {\bibfnamefont {Sevag}\ \bibnamefont
  {Gharibian}}\ and\ \bibinfo {author} {\bibfnamefont {Julia}\ \bibnamefont
  {Kempe}},\ }\bibfield  {title} {\enquote {\bibinfo {title} {Approximation
  algorithms for qma-complete problems},}\ }\href {\doibase 10.1137/110842272}
  {\bibfield  {journal} {\bibinfo  {journal} {SIAM Journal on Computing}\
  }\textbf {\bibinfo {volume} {41}},\ \bibinfo {pages} {1028--1050} (\bibinfo
  {year} {2012})},\ \Eprint
  {http://arxiv.org/abs/https://doi.org/10.1137/110842272}
  {https://doi.org/10.1137/110842272} \BibitemShut {NoStop}%
\bibitem [{\citenamefont {Aharonov}\ \emph {et~al.}(2013)\citenamefont
  {Aharonov}, \citenamefont {Arad},\ and\ \citenamefont
  {Vidick}}]{aharonov2013guest}%
  \BibitemOpen
  \bibfield  {author} {\bibinfo {author} {\bibfnamefont {Dorit}\ \bibnamefont
  {Aharonov}}, \bibinfo {author} {\bibfnamefont {Itai}\ \bibnamefont {Arad}}, \
  and\ \bibinfo {author} {\bibfnamefont {Thomas}\ \bibnamefont {Vidick}},\
  }\bibfield  {title} {\enquote {\bibinfo {title} {Guest column: the quantum
  pcp conjecture},}\ }\href@noop {} {\bibfield  {journal} {\bibinfo  {journal}
  {Acm sigact news}\ }\textbf {\bibinfo {volume} {44}},\ \bibinfo {pages}
  {47--79} (\bibinfo {year} {2013})}\BibitemShut {NoStop}%
\bibitem [{\citenamefont {Brand{\~a}o}\ and\ \citenamefont
  {Harrow}(2016)}]{qPCP}%
  \BibitemOpen
  \bibfield  {author} {\bibinfo {author} {\bibfnamefont {Fernando G. S.~L.}\
  \bibnamefont {Brand{\~a}o}}\ and\ \bibinfo {author} {\bibfnamefont {Aram~W.}\
  \bibnamefont {Harrow}},\ }\bibfield  {title} {\enquote {\bibinfo {title}
  {Product-state approximations to quantum states},}\ }\href {\doibase
  10.1007/s00220-016-2575-1} {\bibfield  {journal} {\bibinfo  {journal}
  {Communications in Mathematical Physics}\ }\textbf {\bibinfo {volume}
  {342}},\ \bibinfo {pages} {47--80} (\bibinfo {year} {2016})}\BibitemShut
  {NoStop}%
\bibitem [{\citenamefont {{Eldar}}\ and\ \citenamefont
  {{Harrow}}(2017)}]{8104078}%
  \BibitemOpen
  \bibfield  {author} {\bibinfo {author} {\bibfnamefont {L.}~\bibnamefont
  {{Eldar}}}\ and\ \bibinfo {author} {\bibfnamefont {A.~W.}\ \bibnamefont
  {{Harrow}}},\ }\bibfield  {title} {\enquote {\bibinfo {title} {Local
  hamiltonians whose ground states are hard to approximate},}\ }in\ \href@noop
  {} {\emph {\bibinfo {booktitle} {2017 IEEE 58th Annual Symposium on
  Foundations of Computer Science (FOCS)}}}\ (\bibinfo {year} {2017})\ pp.\
  \bibinfo {pages} {427--438}\BibitemShut {NoStop}%
\bibitem [{\citenamefont {Nirkhe}\ \emph {et~al.}(2018)\citenamefont {Nirkhe},
  \citenamefont {Vazirani},\ and\ \citenamefont
  {Yuen}}]{nirkhe_et_al:LIPIcs:2018:9095}%
  \BibitemOpen
  \bibfield  {author} {\bibinfo {author} {\bibfnamefont {Chinmay}\ \bibnamefont
  {Nirkhe}}, \bibinfo {author} {\bibfnamefont {Umesh}\ \bibnamefont
  {Vazirani}}, \ and\ \bibinfo {author} {\bibfnamefont {Henry}\ \bibnamefont
  {Yuen}},\ }\bibfield  {title} {\enquote {\bibinfo {title} {{Approximate
  Low-Weight Check Codes and Circuit Lower Bounds for Noisy Ground States}},}\
  }in\ \href {\doibase 10.4230/LIPIcs.ICALP.2018.91} {\emph {\bibinfo
  {booktitle} {45th International Colloquium on Automata, Languages, and
  Programming (ICALP 2018)}}},\ \bibinfo {series} {Leibniz International
  Proceedings in Informatics (LIPIcs)}, Vol.\ \bibinfo {volume} {107},\
  \bibinfo {editor} {edited by\ \bibinfo {editor} {\bibfnamefont {Ioannis}\
  \bibnamefont {Chatzigiannakis}}, \bibinfo {editor} {\bibfnamefont {Christos}\
  \bibnamefont {Kaklamanis}}, \bibinfo {editor} {\bibfnamefont {D{\'a}niel}\
  \bibnamefont {Marx}}, \ and\ \bibinfo {editor} {\bibfnamefont {Donald}\
  \bibnamefont {Sannella}}}\ (\bibinfo  {publisher} {Schloss
  Dagstuhl--Leibniz-Zentrum fuer Informatik},\ \bibinfo {address} {Dagstuhl,
  Germany},\ \bibinfo {year} {2018})\ pp.\ \bibinfo {pages}
  {91:1--91:11}\BibitemShut {NoStop}%
\bibitem [{\citenamefont {Freedman}\ and\ \citenamefont
  {Hastings}(2013)}]{freedman2013quantum}%
  \BibitemOpen
  \bibfield  {author} {\bibinfo {author} {\bibfnamefont {Michael~H}\
  \bibnamefont {Freedman}}\ and\ \bibinfo {author} {\bibfnamefont {Matthew~B}\
  \bibnamefont {Hastings}},\ }\bibfield  {title} {\enquote {\bibinfo {title}
  {Quantum systems on non-$ k $-hyperfinite complexes: A generalization of
  classical statistical mechanics on expander graphs},}\ }\href@noop {}
  {\bibfield  {journal} {\bibinfo  {journal} {arXiv preprint arXiv:1301.1363}\
  } (\bibinfo {year} {2013})}\BibitemShut {NoStop}%
\bibitem [{\citenamefont {Lieb}(1973)}]{lieb1973}%
  \BibitemOpen
  \bibfield  {author} {\bibinfo {author} {\bibfnamefont {Elliott~H.}\
  \bibnamefont {Lieb}},\ }\bibfield  {title} {\enquote {\bibinfo {title} {The
  classical limit of quantum spin systems},}\ }\href
  {https://projecteuclid.org:443/euclid.cmp/1103859040} {\bibfield  {journal}
  {\bibinfo  {journal} {Comm. Math. Phys.}\ }\textbf {\bibinfo {volume} {31}},\
  \bibinfo {pages} {327--340} (\bibinfo {year} {1973})}\BibitemShut {NoStop}%
\bibitem [{\citenamefont {Hsu}\ \emph {et~al.}(2013)\citenamefont {Hsu},
  \citenamefont {Laumann}, \citenamefont {L\"auchli}, \citenamefont
  {Moessner},\ and\ \citenamefont {Sondhi}}]{PhysRevA.87.062334}%
  \BibitemOpen
  \bibfield  {author} {\bibinfo {author} {\bibfnamefont {B.}~\bibnamefont
  {Hsu}}, \bibinfo {author} {\bibfnamefont {C.~R.}\ \bibnamefont {Laumann}},
  \bibinfo {author} {\bibfnamefont {A.~M.}\ \bibnamefont {L\"auchli}}, \bibinfo
  {author} {\bibfnamefont {R.}~\bibnamefont {Moessner}}, \ and\ \bibinfo
  {author} {\bibfnamefont {S.~L.}\ \bibnamefont {Sondhi}},\ }\bibfield  {title}
  {\enquote {\bibinfo {title} {Approximating random quantum optimization
  problems},}\ }\href {\doibase 10.1103/PhysRevA.87.062334} {\bibfield
  {journal} {\bibinfo  {journal} {Phys. Rev. A}\ }\textbf {\bibinfo {volume}
  {87}},\ \bibinfo {pages} {062334} (\bibinfo {year} {2013})}\BibitemShut
  {NoStop}%
\bibitem [{\citenamefont {Bravyi}\ \emph {et~al.}(2019)\citenamefont {Bravyi},
  \citenamefont {Gosset}, \citenamefont {K{\"o}nig},\ and\ \citenamefont
  {Temme}}]{doi:10.1063/1.5085428}%
  \BibitemOpen
  \bibfield  {author} {\bibinfo {author} {\bibfnamefont {Sergey}\ \bibnamefont
  {Bravyi}}, \bibinfo {author} {\bibfnamefont {David}\ \bibnamefont {Gosset}},
  \bibinfo {author} {\bibfnamefont {Robert}\ \bibnamefont {K{\"o}nig}}, \ and\
  \bibinfo {author} {\bibfnamefont {Kristan}\ \bibnamefont {Temme}},\
  }\bibfield  {title} {\enquote {\bibinfo {title} {Approximation algorithms for
  quantum many-body problems},}\ }\href {\doibase 10.1063/1.5085428} {\bibfield
   {journal} {\bibinfo  {journal} {Journal of Mathematical Physics}\ }\textbf
  {\bibinfo {volume} {60}},\ \bibinfo {pages} {032203} (\bibinfo {year}
  {2019})},\ \Eprint {http://arxiv.org/abs/https://doi.org/10.1063/1.5085428}
  {https://doi.org/10.1063/1.5085428} \BibitemShut {NoStop}%
\bibitem [{\citenamefont {Gharibian}\ and\ \citenamefont
  {Parekh}(2019)}]{gharibian2019almost}%
  \BibitemOpen
  \bibfield  {author} {\bibinfo {author} {\bibfnamefont {Sevag}\ \bibnamefont
  {Gharibian}}\ and\ \bibinfo {author} {\bibfnamefont {Ojas}\ \bibnamefont
  {Parekh}},\ }\bibfield  {title} {\enquote {\bibinfo {title} {Almost optimal
  classical approximation algorithms for a quantum generalization of
  max-cut},}\ }\href@noop {} {\bibfield  {journal} {\bibinfo  {journal} {arXiv
  preprint arXiv:1909.08846}\ } (\bibinfo {year} {2019})}\BibitemShut {NoStop}%
\bibitem [{\citenamefont {Sachdev}\ and\ \citenamefont
  {Ye}(1993)}]{Sachdev1993}%
  \BibitemOpen
  \bibfield  {author} {\bibinfo {author} {\bibfnamefont {Subir}\ \bibnamefont
  {Sachdev}}\ and\ \bibinfo {author} {\bibfnamefont {Jinwu}\ \bibnamefont
  {Ye}},\ }\bibfield  {title} {\enquote {\bibinfo {title} {{Gapless spin-fluid
  ground state in a random quantum Heisenberg magnet}},}\ }\href {\doibase
  10.1103/PhysRevLett.70.3339} {\bibfield  {journal} {\bibinfo  {journal}
  {Phys. Rev. Lett.}\ }\textbf {\bibinfo {volume} {70}},\ \bibinfo {pages}
  {3339--3342} (\bibinfo {year} {1993})}\BibitemShut {NoStop}%
\bibitem [{\citenamefont {Kitaev}()}]{KitaevKITP}%
  \BibitemOpen
  \bibfield  {author} {\bibinfo {author} {\bibfnamefont {A.}~\bibnamefont
  {Kitaev}},\ }\bibfield  {title} {\enquote {\bibinfo {title} {{A simple model
  of quantum holography}},}\ }\href
  {http://online.kitp.ucsb.edu/online/entangled15/kitaev/,http:
  //online.kitp.ucsb.edu/online/entangled15/kitaev2/} {\bibinfo  {journal}
  {Talks at KITP, April 7, 2015 and May 27, 2015}\ }\BibitemShut {NoStop}%
\bibitem [{\citenamefont {Maldacena}\ and\ \citenamefont
  {Stanford}(2016)}]{Maldacena2016}%
  \BibitemOpen
\bibfield  {journal} {  }\bibfield  {author} {\bibinfo {author} {\bibfnamefont
  {Juan}\ \bibnamefont {Maldacena}}\ and\ \bibinfo {author} {\bibfnamefont
  {Douglas}\ \bibnamefont {Stanford}},\ }\bibfield  {title} {\enquote {\bibinfo
  {title} {{Remarks on the Sachdev-Ye-Kitaev model}},}\ }\href {\doibase
  10.1103/PhysRevD.94.106002} {\bibfield  {journal} {\bibinfo  {journal} {Phys.
  Rev. D}\ }\textbf {\bibinfo {volume} {94}},\ \bibinfo {pages} {106002}
  (\bibinfo {year} {2016})}\BibitemShut {NoStop}%
\bibitem [{\citenamefont {Parcollet}\ and\ \citenamefont
  {Georges}(1997)}]{Parcollet1997}%
  \BibitemOpen
  \bibfield  {author} {\bibinfo {author} {\bibfnamefont {Olivier}\ \bibnamefont
  {Parcollet}}\ and\ \bibinfo {author} {\bibfnamefont {Antoine}\ \bibnamefont
  {Georges}},\ }\bibfield  {title} {\enquote {\bibinfo {title} {{Transition
  from Overscreening to Underscreening in the Multichannel Kondo Model: Exact
  Solution at Large N}},}\ }\href {\doibase 10.1103/PhysRevLett.79.4665}
  {\bibfield  {journal} {\bibinfo  {journal} {Physical Review Letters}\
  }\textbf {\bibinfo {volume} {79}},\ \bibinfo {pages} {4665--4668} (\bibinfo
  {year} {1997})}\BibitemShut {NoStop}%
\bibitem [{\citenamefont {Gu}\ \emph {et~al.}(2017)\citenamefont {Gu},
  \citenamefont {Lucas},\ and\ \citenamefont {Qi}}]{Gu2017}%
  \BibitemOpen
  \bibfield  {author} {\bibinfo {author} {\bibfnamefont {Yingfei}\ \bibnamefont
  {Gu}}, \bibinfo {author} {\bibfnamefont {Andrew}\ \bibnamefont {Lucas}}, \
  and\ \bibinfo {author} {\bibfnamefont {Xiao-Liang}\ \bibnamefont {Qi}},\
  }\bibfield  {title} {\enquote {\bibinfo {title} {{Spread of entanglement in a
  Sachdev-Ye-Kitaev chain}},}\ }\href@noop {} {\  (\bibinfo {year} {2017})},\
  \Eprint {http://arxiv.org/abs/1708.00871} {arXiv:1708.00871 [hep-th]}
  \BibitemShut {NoStop}%
%%CITATION = ARXIV:1708.00871;%%
\bibitem [{\citenamefont {Banerjee}\ and\ \citenamefont
  {Altman}(2017)}]{BanerjeeAltman2016}%
  \BibitemOpen
  \bibfield  {author} {\bibinfo {author} {\bibfnamefont {Sumilan}\ \bibnamefont
  {Banerjee}}\ and\ \bibinfo {author} {\bibfnamefont {Ehud}\ \bibnamefont
  {Altman}},\ }\bibfield  {title} {\enquote {\bibinfo {title} {Solvable model
  for a dynamical quantum phase transition from fast to slow scrambling},}\
  }\href {\doibase 10.1103/PhysRevB.95.134302} {\bibfield  {journal} {\bibinfo
  {journal} {Phys. Rev. B}\ }\textbf {\bibinfo {volume} {95}},\ \bibinfo
  {pages} {134302} (\bibinfo {year} {2017})}\BibitemShut {NoStop}%
\bibitem [{\citenamefont {Jian}\ and\ \citenamefont {Yao}(2017)}]{SJian2017b}%
  \BibitemOpen
  \bibfield  {author} {\bibinfo {author} {\bibfnamefont {Shao-Kai}\
  \bibnamefont {Jian}}\ and\ \bibinfo {author} {\bibfnamefont {Hong}\
  \bibnamefont {Yao}},\ }\bibfield  {title} {\enquote {\bibinfo {title}
  {{Solvable Sachdev-Ye-Kitaev Models in Higher Dimensions: From Diffusion to
  Many-Body Localization}},}\ }\href {\doibase 10.1103/PhysRevLett.119.206602}
  {\bibfield  {journal} {\bibinfo  {journal} {Phys. Rev. Lett.}\ }\textbf
  {\bibinfo {volume} {119}},\ \bibinfo {pages} {206602} (\bibinfo {year}
  {2017})}\BibitemShut {NoStop}%
\bibitem [{\citenamefont {Song}\ \emph {et~al.}(2017)\citenamefont {Song},
  \citenamefont {Jian},\ and\ \citenamefont {Balents}}]{Song2017}%
  \BibitemOpen
  \bibfield  {author} {\bibinfo {author} {\bibfnamefont {Xue-Yang}\
  \bibnamefont {Song}}, \bibinfo {author} {\bibfnamefont {Chao-Ming}\
  \bibnamefont {Jian}}, \ and\ \bibinfo {author} {\bibfnamefont {Leon}\
  \bibnamefont {Balents}},\ }\bibfield  {title} {\enquote {\bibinfo {title}
  {{Strongly Correlated Metal Built from Sachdev-Ye-Kitaev Models}},}\ }\href
  {\doibase 10.1103/PhysRevLett.119.216601} {\bibfield  {journal} {\bibinfo
  {journal} {Phys. Rev. Lett.}\ }\textbf {\bibinfo {volume} {119}},\ \bibinfo
  {pages} {216601} (\bibinfo {year} {2017})}\BibitemShut {NoStop}%
\bibitem [{\citenamefont {Davison}\ \emph {et~al.}(2017)\citenamefont
  {Davison}, \citenamefont {Fu}, \citenamefont {Georges}, \citenamefont {Gu},
  \citenamefont {Jensen},\ and\ \citenamefont {Sachdev}}]{Davison2017}%
  \BibitemOpen
  \bibfield  {author} {\bibinfo {author} {\bibfnamefont {Richard~A.}\
  \bibnamefont {Davison}}, \bibinfo {author} {\bibfnamefont {Wenbo}\
  \bibnamefont {Fu}}, \bibinfo {author} {\bibfnamefont {Antoine}\ \bibnamefont
  {Georges}}, \bibinfo {author} {\bibfnamefont {Yingfei}\ \bibnamefont {Gu}},
  \bibinfo {author} {\bibfnamefont {Kristan}\ \bibnamefont {Jensen}}, \ and\
  \bibinfo {author} {\bibfnamefont {Subir}\ \bibnamefont {Sachdev}},\
  }\bibfield  {title} {\enquote {\bibinfo {title} {Thermoelectric transport in
  disordered metals without quasiparticles: The sachdev-ye-kitaev models and
  holography},}\ }\href {\doibase 10.1103/PhysRevB.95.155131} {\bibfield
  {journal} {\bibinfo  {journal} {Phys. Rev. B}\ }\textbf {\bibinfo {volume}
  {95}},\ \bibinfo {pages} {155131} (\bibinfo {year} {2017})}\BibitemShut
  {NoStop}%
\bibitem [{\citenamefont {Haldar}\ \emph {et~al.}(2018)\citenamefont {Haldar},
  \citenamefont {Banerjee},\ and\ \citenamefont {Shenoy}}]{Arijit2017}%
  \BibitemOpen
  \bibfield  {author} {\bibinfo {author} {\bibfnamefont {Arijit}\ \bibnamefont
  {Haldar}}, \bibinfo {author} {\bibfnamefont {Sumilan}\ \bibnamefont
  {Banerjee}}, \ and\ \bibinfo {author} {\bibfnamefont {Vijay~B.}\ \bibnamefont
  {Shenoy}},\ }\bibfield  {title} {\enquote {\bibinfo {title}
  {{Higher-dimensional Sachdev-Ye-Kitaev non-Fermi liquids at Lifshitz
  transitions}},}\ }\href {\doibase 10.1103/PhysRevB.97.241106} {\bibfield
  {journal} {\bibinfo  {journal} {Phys. Rev. B}\ }\textbf {\bibinfo {volume}
  {97}},\ \bibinfo {pages} {241106} (\bibinfo {year} {2018})}\BibitemShut
  {NoStop}%
\bibitem [{\citenamefont {Patel}\ \emph {et~al.}(2018)\citenamefont {Patel},
  \citenamefont {McGreevy}, \citenamefont {Arovas},\ and\ \citenamefont
  {Sachdev}}]{Patel2018}%
  \BibitemOpen
  \bibfield  {author} {\bibinfo {author} {\bibfnamefont {Aavishkar~A.}\
  \bibnamefont {Patel}}, \bibinfo {author} {\bibfnamefont {John}\ \bibnamefont
  {McGreevy}}, \bibinfo {author} {\bibfnamefont {Daniel~P.}\ \bibnamefont
  {Arovas}}, \ and\ \bibinfo {author} {\bibfnamefont {Subir}\ \bibnamefont
  {Sachdev}},\ }\bibfield  {title} {\enquote {\bibinfo {title}
  {Magnetotransport in a model of a disordered strange metal},}\ }\href
  {\doibase 10.1103/PhysRevX.8.021049} {\bibfield  {journal} {\bibinfo
  {journal} {Phys. Rev. X}\ }\textbf {\bibinfo {volume} {8}},\ \bibinfo {pages}
  {021049} (\bibinfo {year} {2018})}\BibitemShut {NoStop}%
\bibitem [{\citenamefont {Chowdhury}\ \emph {et~al.}(2018)\citenamefont
  {Chowdhury}, \citenamefont {Werman}, \citenamefont {Berg},\ and\
  \citenamefont {Senthil}}]{Chowdhury2018}%
  \BibitemOpen
  \bibfield  {author} {\bibinfo {author} {\bibfnamefont {Debanjan}\
  \bibnamefont {Chowdhury}}, \bibinfo {author} {\bibfnamefont {Yochai}\
  \bibnamefont {Werman}}, \bibinfo {author} {\bibfnamefont {Erez}\ \bibnamefont
  {Berg}}, \ and\ \bibinfo {author} {\bibfnamefont {T.}~\bibnamefont
  {Senthil}},\ }\bibfield  {title} {\enquote {\bibinfo {title} {Translationally
  invariant non-fermi-liquid metals with critical fermi surfaces: Solvable
  models},}\ }\href {\doibase 10.1103/PhysRevX.8.031024} {\bibfield  {journal}
  {\bibinfo  {journal} {Phys. Rev. X}\ }\textbf {\bibinfo {volume} {8}},\
  \bibinfo {pages} {031024} (\bibinfo {year} {2018})}\BibitemShut {NoStop}%
\bibitem [{\citenamefont {Haldar}\ and\ \citenamefont
  {Shenoy}(2018)}]{Haldar2018PRB}%
  \BibitemOpen
  \bibfield  {author} {\bibinfo {author} {\bibfnamefont {Arijit}\ \bibnamefont
  {Haldar}}\ and\ \bibinfo {author} {\bibfnamefont {Vijay~B.}\ \bibnamefont
  {Shenoy}},\ }\bibfield  {title} {\enquote {\bibinfo {title} {Strange
  half-metals and mott insulators in sachdev-ye-kitaev models},}\ }\href
  {\doibase 10.1103/PhysRevB.98.165135} {\bibfield  {journal} {\bibinfo
  {journal} {Phys. Rev. B}\ }\textbf {\bibinfo {volume} {98}},\ \bibinfo
  {pages} {165135} (\bibinfo {year} {2018})}\BibitemShut {NoStop}%
\bibitem [{\citenamefont {{Haldar}}\ \emph {et~al.}(2020)\citenamefont
  {{Haldar}}, \citenamefont {{Bera}},\ and\ \citenamefont
  {{Banerjee}}}]{Haldar2020arXivRenyi}%
  \BibitemOpen
  \bibfield  {author} {\bibinfo {author} {\bibfnamefont {Arijit}\ \bibnamefont
  {{Haldar}}}, \bibinfo {author} {\bibfnamefont {Surajit}\ \bibnamefont
  {{Bera}}}, \ and\ \bibinfo {author} {\bibfnamefont {Sumilan}\ \bibnamefont
  {{Banerjee}}},\ }\bibfield  {title} {\enquote {\bibinfo {title} {{R\'{e}nyi
  entanglement entropy of Fermi liquids and non-Fermi liquids:
  Sachdev-Ye-Kitaev model and dynamical mean field theories}},}\ }\href@noop {}
  {\bibfield  {journal} {\bibinfo  {journal} {arXiv e-prints}\ ,\ \bibinfo
  {eid} {arXiv:2004.04751}} (\bibinfo {year} {2020})},\ \Eprint
  {http://arxiv.org/abs/2004.04751} {arXiv:2004.04751 [cond-mat.str-el]}
  \BibitemShut {NoStop}%
\bibitem [{\citenamefont {Scaffidi}\ and\ \citenamefont
  {Altman}(2019)}]{Scaffidi2019PRB}%
  \BibitemOpen
  \bibfield  {author} {\bibinfo {author} {\bibfnamefont {Thomas}\ \bibnamefont
  {Scaffidi}}\ and\ \bibinfo {author} {\bibfnamefont {Ehud}\ \bibnamefont
  {Altman}},\ }\bibfield  {title} {\enquote {\bibinfo {title} {Chaos in a
  classical limit of the sachdev-ye-kitaev model},}\ }\href {\doibase
  10.1103/PhysRevB.100.155128} {\bibfield  {journal} {\bibinfo  {journal}
  {Phys. Rev. B}\ }\textbf {\bibinfo {volume} {100}},\ \bibinfo {pages}
  {155128} (\bibinfo {year} {2019})}\BibitemShut {NoStop}%
\bibitem [{\citenamefont {Parker}\ \emph {et~al.}(2019)\citenamefont {Parker},
  \citenamefont {Cao}, \citenamefont {Avdoshkin}, \citenamefont {Scaffidi},\
  and\ \citenamefont {Altman}}]{Scaffidi2019PRX}%
  \BibitemOpen
  \bibfield  {author} {\bibinfo {author} {\bibfnamefont {Daniel~E.}\
  \bibnamefont {Parker}}, \bibinfo {author} {\bibfnamefont {Xiangyu}\
  \bibnamefont {Cao}}, \bibinfo {author} {\bibfnamefont {Alexander}\
  \bibnamefont {Avdoshkin}}, \bibinfo {author} {\bibfnamefont {Thomas}\
  \bibnamefont {Scaffidi}}, \ and\ \bibinfo {author} {\bibfnamefont {Ehud}\
  \bibnamefont {Altman}},\ }\bibfield  {title} {\enquote {\bibinfo {title} {A
  universal operator growth hypothesis},}\ }\href {\doibase
  10.1103/PhysRevX.9.041017} {\bibfield  {journal} {\bibinfo  {journal} {Phys.
  Rev. X}\ }\textbf {\bibinfo {volume} {9}},\ \bibinfo {pages} {041017}
  (\bibinfo {year} {2019})}\BibitemShut {NoStop}%
\bibitem [{\citenamefont {Eberlein}\ \emph {et~al.}(2017)\citenamefont
  {Eberlein}, \citenamefont {Kasper}, \citenamefont {Sachdev},\ and\
  \citenamefont {Steinberg}}]{Eberlein2017}%
  \BibitemOpen
  \bibfield  {author} {\bibinfo {author} {\bibfnamefont {Andreas}\ \bibnamefont
  {Eberlein}}, \bibinfo {author} {\bibfnamefont {Valentin}\ \bibnamefont
  {Kasper}}, \bibinfo {author} {\bibfnamefont {Subir}\ \bibnamefont {Sachdev}},
  \ and\ \bibinfo {author} {\bibfnamefont {Julia}\ \bibnamefont {Steinberg}},\
  }\bibfield  {title} {\enquote {\bibinfo {title} {Quantum quench of the
  sachdev-ye-kitaev model},}\ }\href {\doibase 10.1103/PhysRevB.96.205123}
  {\bibfield  {journal} {\bibinfo  {journal} {Phys. Rev. B}\ }\textbf {\bibinfo
  {volume} {96}},\ \bibinfo {pages} {205123} (\bibinfo {year}
  {2017})}\BibitemShut {NoStop}%
\bibitem [{\citenamefont {Haldar}\ \emph {et~al.}(2020)\citenamefont {Haldar},
  \citenamefont {Haldar}, \citenamefont {Bera}, \citenamefont {Mandal},\ and\
  \citenamefont {Banerjee}}]{Haldar2020}%
  \BibitemOpen
  \bibfield  {author} {\bibinfo {author} {\bibfnamefont {Arijit}\ \bibnamefont
  {Haldar}}, \bibinfo {author} {\bibfnamefont {Prosenjit}\ \bibnamefont
  {Haldar}}, \bibinfo {author} {\bibfnamefont {Surajit}\ \bibnamefont {Bera}},
  \bibinfo {author} {\bibfnamefont {Ipsita}\ \bibnamefont {Mandal}}, \ and\
  \bibinfo {author} {\bibfnamefont {Sumilan}\ \bibnamefont {Banerjee}},\
  }\bibfield  {title} {\enquote {\bibinfo {title} {Quench, thermalization, and
  residual entropy across a non-fermi liquid to fermi liquid transition},}\
  }\href {\doibase 10.1103/PhysRevResearch.2.013307} {\bibfield  {journal}
  {\bibinfo  {journal} {Phys. Rev. Research}\ }\textbf {\bibinfo {volume}
  {2}},\ \bibinfo {pages} {013307} (\bibinfo {year} {2020})}\BibitemShut
  {NoStop}%
\bibitem [{\citenamefont {Almheiri}\ \emph {et~al.}(2019)\citenamefont
  {Almheiri}, \citenamefont {Milekhin},\ and\ \citenamefont
  {Swingle}}]{Almheiri2019}%
  \BibitemOpen
  \bibfield  {author} {\bibinfo {author} {\bibfnamefont {Ahmed}\ \bibnamefont
  {Almheiri}}, \bibinfo {author} {\bibfnamefont {Alexey}\ \bibnamefont
  {Milekhin}}, \ and\ \bibinfo {author} {\bibfnamefont {Brian}\ \bibnamefont
  {Swingle}},\ }\bibfield  {title} {\enquote {\bibinfo {title} {{Universal
  Constraints on Energy Flow and SYK Thermalization}},}\ }\href@noop {} {\
  (\bibinfo {year} {2019})},\ \Eprint {http://arxiv.org/abs/1912.04912}
  {arXiv:1912.04912 [hep-th]} \BibitemShut {NoStop}%
\bibitem [{\citenamefont {Sachdev}(2010)}]{Sachdev2010}%
  \BibitemOpen
  \bibfield  {author} {\bibinfo {author} {\bibfnamefont {Subir}\ \bibnamefont
  {Sachdev}},\ }\bibfield  {title} {\enquote {\bibinfo {title} {{Holographic
  Metals and the Fractionalized Fermi Liquid}},}\ }\href {\doibase
  10.1103/PhysRevLett.105.151602} {\bibfield  {journal} {\bibinfo  {journal}
  {Phys. Rev. Lett.}\ }\textbf {\bibinfo {volume} {105}},\ \bibinfo {pages}
  {151602} (\bibinfo {year} {2010})}\BibitemShut {NoStop}%
\bibitem [{\citenamefont {Sachdev}(2015)}]{Sachdev2015}%
  \BibitemOpen
  \bibfield  {author} {\bibinfo {author} {\bibfnamefont {Subir}\ \bibnamefont
  {Sachdev}},\ }\bibfield  {title} {\enquote {\bibinfo {title}
  {{Bekenstein-hawking entropy and strange metals}},}\ }\href {\doibase
  10.1103/PhysRevX.5.041025} {\bibfield  {journal} {\bibinfo  {journal}
  {Physical Review X}\ }\textbf {\bibinfo {volume} {5}},\ \bibinfo {pages}
  {1--13} (\bibinfo {year} {2015})}\BibitemShut {NoStop}%
\bibitem [{\citenamefont {Maldacena}\ and\ \citenamefont
  {Qi}(2018)}]{Maldacena2018lmt}%
  \BibitemOpen
  \bibfield  {author} {\bibinfo {author} {\bibfnamefont {Juan}\ \bibnamefont
  {Maldacena}}\ and\ \bibinfo {author} {\bibfnamefont {Xiao-Liang}\
  \bibnamefont {Qi}},\ }\bibfield  {title} {\enquote {\bibinfo {title}
  {{Eternal traversable wormhole}},}\ }\href@noop {} {\  (\bibinfo {year}
  {2018})},\ \Eprint {http://arxiv.org/abs/1804.00491} {arXiv:1804.00491
  [hep-th]} \BibitemShut {NoStop}%
\bibitem [{\citenamefont {Kitaev}\ and\ \citenamefont
  {Suh}(2018)}]{kitaev2018soft}%
  \BibitemOpen
  \bibfield  {author} {\bibinfo {author} {\bibfnamefont {Alexei}\ \bibnamefont
  {Kitaev}}\ and\ \bibinfo {author} {\bibfnamefont {S~Josephine}\ \bibnamefont
  {Suh}},\ }\bibfield  {title} {\enquote {\bibinfo {title} {The soft mode in
  the sachdev-ye-kitaev model and its gravity dual},}\ }\href@noop {}
  {\bibfield  {journal} {\bibinfo  {journal} {Journal of High Energy Physics}\
  }\textbf {\bibinfo {volume} {2018}},\ \bibinfo {pages} {183} (\bibinfo {year}
  {2018})}\BibitemShut {NoStop}%
\bibitem [{\citenamefont {Qi}\ and\ \citenamefont
  {Zhang}(2020)}]{qi2020coupled}%
  \BibitemOpen
  \bibfield  {author} {\bibinfo {author} {\bibfnamefont {Xiao-Liang}\
  \bibnamefont {Qi}}\ and\ \bibinfo {author} {\bibfnamefont {Pengfei}\
  \bibnamefont {Zhang}},\ }\bibfield  {title} {\enquote {\bibinfo {title} {The
  coupled syk model at finite temperature},}\ }\href@noop {} {\bibfield
  {journal} {\bibinfo  {journal} {Journal of High Energy Physics}\ }\textbf
  {\bibinfo {volume} {2020}},\ \bibinfo {pages} {1--14} (\bibinfo {year}
  {2020})}\BibitemShut {NoStop}%
\bibitem [{\citenamefont {Gao}\ \emph {et~al.}(2017)\citenamefont {Gao},
  \citenamefont {Jafferis},\ and\ \citenamefont {Wall}}]{gao2017traversable}%
  \BibitemOpen
  \bibfield  {author} {\bibinfo {author} {\bibfnamefont {Ping}\ \bibnamefont
  {Gao}}, \bibinfo {author} {\bibfnamefont {Daniel~Louis}\ \bibnamefont
  {Jafferis}}, \ and\ \bibinfo {author} {\bibfnamefont {Aron~C}\ \bibnamefont
  {Wall}},\ }\bibfield  {title} {\enquote {\bibinfo {title} {Traversable
  wormholes via a double trace deformation},}\ }\href@noop {} {\bibfield
  {journal} {\bibinfo  {journal} {Journal of High Energy Physics}\ }\textbf
  {\bibinfo {volume} {2017}},\ \bibinfo {pages} {151} (\bibinfo {year}
  {2017})}\BibitemShut {NoStop}%
\bibitem [{\citenamefont {Maldacena}\ \emph {et~al.}(2017)\citenamefont
  {Maldacena}, \citenamefont {Stanford},\ and\ \citenamefont
  {Yang}}]{maldacena2017diving}%
  \BibitemOpen
  \bibfield  {author} {\bibinfo {author} {\bibfnamefont {Juan}\ \bibnamefont
  {Maldacena}}, \bibinfo {author} {\bibfnamefont {Douglas}\ \bibnamefont
  {Stanford}}, \ and\ \bibinfo {author} {\bibfnamefont {Zhenbin}\ \bibnamefont
  {Yang}},\ }\bibfield  {title} {\enquote {\bibinfo {title} {Diving into
  traversable wormholes},}\ }\href@noop {} {\bibfield  {journal} {\bibinfo
  {journal} {Fortschritte der Physik}\ }\textbf {\bibinfo {volume} {65}},\
  \bibinfo {pages} {1700034} (\bibinfo {year} {2017})}\BibitemShut {NoStop}%
\bibitem [{\citenamefont {Kourkoulou}\ and\ \citenamefont
  {Maldacena}(2017)}]{Kourkoulou2017}%
  \BibitemOpen
  \bibfield  {author} {\bibinfo {author} {\bibfnamefont {Ioanna}\ \bibnamefont
  {Kourkoulou}}\ and\ \bibinfo {author} {\bibfnamefont {Juan}\ \bibnamefont
  {Maldacena}},\ }\bibfield  {title} {\enquote {\bibinfo {title} {{Pure states
  in the SYK model and nearly-$AdS_2$ gravity}},}\ }\href@noop {} {\  (\bibinfo
  {year} {2017})},\ \Eprint {http://arxiv.org/abs/1707.02325} {arXiv:1707.02325
  [hep-th]} \BibitemShut {NoStop}%
%%CITATION = ARXIV:1707.02325;%%
\bibitem [{\citenamefont {Zhang}(2020)}]{zhang2020entanglement}%
  \BibitemOpen
  \bibfield  {author} {\bibinfo {author} {\bibfnamefont {Pengfei}\ \bibnamefont
  {Zhang}},\ }\bibfield  {title} {\enquote {\bibinfo {title} {Entanglement
  entropy and its quench dynamics for pure states of the sachdev-ye-kitaev
  model},}\ }\href {\doibase 10.1007/JHEP06(2020)143} {\bibfield  {journal}
  {\bibinfo  {journal} {Journal of High Energy Physics}\ }\textbf {\bibinfo
  {volume} {2020}},\ \bibinfo {pages} {143} (\bibinfo {year}
  {2020})}\BibitemShut {NoStop}%
\bibitem [{Note1()}]{Note1}%
  \BibitemOpen
  \bibinfo {note} {E.g., when $q=4$, we have
  $J_{i_1,i_2;j_1,j_2}=-J_{i_2,i_1;j_1,j_2}=J_{i_2,i_1;j_2,j_1}$ due to fermion
  anti-commutation relations, and $J_{i_1,i_2;j_1,j_2}=J_{j_2,j_1;i_2,i_1}^*$
  for maintaining hermiticity.}\BibitemShut {Stop}%
\bibitem [{Note2()}]{Note2}%
  \BibitemOpen
  \bibinfo {note} {Strictly speaking, particle-hole symmetry is only present in
  the $N \to \infty $ limit, but all of our results are obtained in that limit
  anyway.}\BibitemShut {Stop}%
\bibitem [{\citenamefont {Blinder}(2019)}]{BLINDER20191}%
  \BibitemOpen
  \bibfield  {author} {\bibinfo {author} {\bibfnamefont {S.M.}\ \bibnamefont
  {Blinder}},\ }\bibfield  {title} {\enquote {\bibinfo {title} {Chapter 1 -
  introduction to the hartree-fock method},}\ }in\ \href {\doibase
  https://doi.org/10.1016/B978-0-12-813651-5.00001-2} {\emph {\bibinfo
  {booktitle} {Mathematical Physics in Theoretical Chemistry}}},\ \bibinfo
  {series and number} {Developments in Physical \& Theoretical Chemistry},\
  \bibinfo {editor} {edited by\ \bibinfo {editor} {\bibfnamefont {S.M.}\
  \bibnamefont {Blinder}}\ and\ \bibinfo {editor} {\bibfnamefont {J.E.}\
  \bibnamefont {House}}}\ (\bibinfo  {publisher} {Elsevier},\ \bibinfo {year}
  {2019})\ pp.\ \bibinfo {pages} {1 -- 30}\BibitemShut {NoStop}%
\bibitem [{\citenamefont {Bartlett}\ and\ \citenamefont {Noga}(1988)}]{vCC1}%
  \BibitemOpen
  \bibfield  {author} {\bibinfo {author} {\bibfnamefont {Rodney~J.}\
  \bibnamefont {Bartlett}}\ and\ \bibinfo {author} {\bibfnamefont {Jozef}\
  \bibnamefont {Noga}},\ }\bibfield  {title} {\enquote {\bibinfo {title} {The
  expectation value coupled-cluster method and analytical energy
  derivatives},}\ }\href {\doibase
  https://doi.org/10.1016/0009-2614(88)80392-0} {\bibfield  {journal} {\bibinfo
   {journal} {Chemical Physics Letters}\ }\textbf {\bibinfo {volume} {150}},\
  \bibinfo {pages} {29 -- 36} (\bibinfo {year} {1988})}\BibitemShut {NoStop}%
\bibitem [{\citenamefont {Szalay}\ \emph {et~al.}(1995)\citenamefont {Szalay},
  \citenamefont {Nooijen},\ and\ \citenamefont {Bartlett}}]{vCC2}%
  \BibitemOpen
  \bibfield  {author} {\bibinfo {author} {\bibfnamefont {P{\'e}ter~G.}\
  \bibnamefont {Szalay}}, \bibinfo {author} {\bibfnamefont {Marcel}\
  \bibnamefont {Nooijen}}, \ and\ \bibinfo {author} {\bibfnamefont {Rodney~J.}\
  \bibnamefont {Bartlett}},\ }\bibfield  {title} {\enquote {\bibinfo {title}
  {Alternative ans{\"a}tze in single reference coupled‐cluster theory. iii. a
  critical analysis of different methods},}\ }\href {\doibase 10.1063/1.469641}
  {\bibfield  {journal} {\bibinfo  {journal} {The Journal of Chemical Physics}\
  }\textbf {\bibinfo {volume} {103}},\ \bibinfo {pages} {281--298} (\bibinfo
  {year} {1995})},\ \Eprint
  {http://arxiv.org/abs/https://doi.org/10.1063/1.469641}
  {https://doi.org/10.1063/1.469641} \BibitemShut {NoStop}%
\bibitem [{\citenamefont {Van~Voorhis}\ and\ \citenamefont
  {Head-Gordon}(2000)}]{vCC3}%
  \BibitemOpen
  \bibfield  {author} {\bibinfo {author} {\bibfnamefont {Troy}\ \bibnamefont
  {Van~Voorhis}}\ and\ \bibinfo {author} {\bibfnamefont {Martin}\ \bibnamefont
  {Head-Gordon}},\ }\bibfield  {title} {\enquote {\bibinfo {title} {Benchmark
  variational coupled cluster doubles results},}\ }\href {\doibase
  10.1063/1.1319643} {\bibfield  {journal} {\bibinfo  {journal} {The Journal of
  Chemical Physics}\ }\textbf {\bibinfo {volume} {113}},\ \bibinfo {pages}
  {8873--8879} (\bibinfo {year} {2000})},\ \Eprint
  {http://arxiv.org/abs/https://doi.org/10.1063/1.1319643}
  {https://doi.org/10.1063/1.1319643} \BibitemShut {NoStop}%
\bibitem [{\citenamefont {Bastianello}\ and\ \citenamefont
  {Sotiriadis}(2016)}]{Bastianello2016NPB}%
  \BibitemOpen
  \bibfield  {author} {\bibinfo {author} {\bibfnamefont {Alvise}\ \bibnamefont
  {Bastianello}}\ and\ \bibinfo {author} {\bibfnamefont {Spyros}\ \bibnamefont
  {Sotiriadis}},\ }\bibfield  {title} {\enquote {\bibinfo {title} {Cluster
  expansion for ground states of local hamiltonians},}\ }\href {\doibase
  https://doi.org/10.1016/j.nuclphysb.2016.06.021} {\bibfield  {journal}
  {\bibinfo  {journal} {Nuclear Physics B}\ }\textbf {\bibinfo {volume}
  {909}},\ \bibinfo {pages} {1020 -- 1078} (\bibinfo {year}
  {2016})}\BibitemShut {NoStop}%
\bibitem [{\citenamefont {Coester}\ and\ \citenamefont
  {K{\"u}mmel}(1960)}]{coester1960short}%
  \BibitemOpen
  \bibfield  {author} {\bibinfo {author} {\bibfnamefont {Fritz}\ \bibnamefont
  {Coester}}\ and\ \bibinfo {author} {\bibfnamefont {Hermann}\ \bibnamefont
  {K{\"u}mmel}},\ }\bibfield  {title} {\enquote {\bibinfo {title} {Short-range
  correlations in nuclear wave functions},}\ }\href@noop {} {\bibfield
  {journal} {\bibinfo  {journal} {Nuclear Physics}\ }\textbf {\bibinfo {volume}
  {17}},\ \bibinfo {pages} {477--485} (\bibinfo {year} {1960})}\BibitemShut
  {NoStop}%
\bibitem [{\citenamefont
  {{\v{C}}{\'\i}{\v{z}}ek}(1966)}]{vcivzek1966correlation}%
  \BibitemOpen
  \bibfield  {author} {\bibinfo {author} {\bibfnamefont {Ji{\v{r}}{\'\i}}\
  \bibnamefont {{\v{C}}{\'\i}{\v{z}}ek}},\ }\bibfield  {title} {\enquote
  {\bibinfo {title} {On the correlation problem in atomic and molecular
  systems. calculation of wavefunction components in ursell-type expansion
  using quantum-field theoretical methods},}\ }\href@noop {} {\bibfield
  {journal} {\bibinfo  {journal} {The Journal of Chemical Physics}\ }\textbf
  {\bibinfo {volume} {45}},\ \bibinfo {pages} {4256--4266} (\bibinfo {year}
  {1966})}\BibitemShut {NoStop}%
\bibitem [{Note3()}]{Note3}%
  \BibitemOpen
  \bibinfo {note} {In usual applications of coupled cluster theory, the
  partitioning of the system is decided by the Hartee-Fock method, which
  separates orbitals that are occupied in the Hartree-Fock state from the
  others. In that setting, the left-right asymmetry is natural, and measures
  how many particle-hole excitations from the reference Hartree-Fock state are
  created in order to accommodate the interaction terms in the Hamiltonian. By
  contrast, in our case the partitioning between left and right is artificial
  and the fact that $|\psi (a_{min})$$\rangle $ is unbalanced is an artefact of
  the technique that should not be present in the true eigenstates of
  $H_{SYK}$. However, when studying $H_{pSYK}$, this balance is actually
  physical and is a manifestation of spontaneous symmetry
  breaking.}\BibitemShut {Stop}%
\bibitem [{\citenamefont {Fu}\ and\ \citenamefont {Sachdev}(2016)}]{Fu2016}%
  \BibitemOpen
  \bibfield  {author} {\bibinfo {author} {\bibfnamefont {Wenbo}\ \bibnamefont
  {Fu}}\ and\ \bibinfo {author} {\bibfnamefont {Subir}\ \bibnamefont
  {Sachdev}},\ }\bibfield  {title} {\enquote {\bibinfo {title} {{Numerical
  study of fermion and boson models with infinite-range random
  interactions}},}\ }\href {\doibase 10.1103/PhysRevB.94.035135} {\bibfield
  {journal} {\bibinfo  {journal} {Phys. Rev. B}\ }\textbf {\bibinfo {volume}
  {94}},\ \bibinfo {pages} {035135} (\bibinfo {year} {2016})}\BibitemShut
  {NoStop}%
\bibitem [{\citenamefont {Liu}\ \emph {et~al.}(2018)\citenamefont {Liu},
  \citenamefont {Chen},\ and\ \citenamefont {Balents}}]{BalentsPRB2018}%
  \BibitemOpen
  \bibfield  {author} {\bibinfo {author} {\bibfnamefont {Chunxiao}\
  \bibnamefont {Liu}}, \bibinfo {author} {\bibfnamefont {Xiao}\ \bibnamefont
  {Chen}}, \ and\ \bibinfo {author} {\bibfnamefont {Leon}\ \bibnamefont
  {Balents}},\ }\bibfield  {title} {\enquote {\bibinfo {title} {Quantum
  entanglement of the sachdev-ye-kitaev models},}\ }\href {\doibase
  10.1103/PhysRevB.97.245126} {\bibfield  {journal} {\bibinfo  {journal} {Phys.
  Rev. B}\ }\textbf {\bibinfo {volume} {97}},\ \bibinfo {pages} {245126}
  (\bibinfo {year} {2018})}\BibitemShut {NoStop}%
\bibitem [{\citenamefont {Goel}\ \emph {et~al.}(2019)\citenamefont {Goel},
  \citenamefont {Lam}, \citenamefont {Turiaci},\ and\ \citenamefont
  {Verlinde}}]{GoelJHEP2019}%
  \BibitemOpen
  \bibfield  {author} {\bibinfo {author} {\bibfnamefont {Akash}\ \bibnamefont
  {Goel}}, \bibinfo {author} {\bibfnamefont {Ho~Tat}\ \bibnamefont {Lam}},
  \bibinfo {author} {\bibfnamefont {Gustavo~J.}\ \bibnamefont {Turiaci}}, \
  and\ \bibinfo {author} {\bibfnamefont {Herman}\ \bibnamefont {Verlinde}},\
  }\bibfield  {title} {\enquote {\bibinfo {title} {Expanding the black hole
  interior: partially entangled thermal states in syk},}\ }\href {\doibase
  10.1007/JHEP02(2019)156} {\bibfield  {journal} {\bibinfo  {journal} {Journal
  of High Energy Physics}\ }\textbf {\bibinfo {volume} {2019}},\ \bibinfo
  {pages} {156} (\bibinfo {year} {2019})}\BibitemShut {NoStop}%
\bibitem [{\citenamefont {Huang}\ and\ \citenamefont {Gu}(2019)}]{GuPRD2019}%
  \BibitemOpen
  \bibfield  {author} {\bibinfo {author} {\bibfnamefont {Yichen}\ \bibnamefont
  {Huang}}\ and\ \bibinfo {author} {\bibfnamefont {Yingfei}\ \bibnamefont
  {Gu}},\ }\bibfield  {title} {\enquote {\bibinfo {title} {Eigenstate
  entanglement in the sachdev-ye-kitaev model},}\ }\href {\doibase
  10.1103/PhysRevD.100.041901} {\bibfield  {journal} {\bibinfo  {journal}
  {Phys. Rev. D}\ }\textbf {\bibinfo {volume} {100}},\ \bibinfo {pages}
  {041901} (\bibinfo {year} {2019})}\BibitemShut {NoStop}%
\bibitem [{\citenamefont {O’Malley}\ \emph {et~al.}(2016)\citenamefont
  {O’Malley}, \citenamefont {Babbush}, \citenamefont {Kivlichan},
  \citenamefont {Romero}, \citenamefont {McClean}, \citenamefont {Barends},
  \citenamefont {Kelly}, \citenamefont {Roushan}, \citenamefont {Tranter},
  \citenamefont {Ding} \emph {et~al.}}]{o2016scalable}%
  \BibitemOpen
  \bibfield  {author} {\bibinfo {author} {\bibfnamefont {Peter~JJ}\
  \bibnamefont {O’Malley}}, \bibinfo {author} {\bibfnamefont {Ryan}\
  \bibnamefont {Babbush}}, \bibinfo {author} {\bibfnamefont {Ian~D}\
  \bibnamefont {Kivlichan}}, \bibinfo {author} {\bibfnamefont {Jonathan}\
  \bibnamefont {Romero}}, \bibinfo {author} {\bibfnamefont {Jarrod~R}\
  \bibnamefont {McClean}}, \bibinfo {author} {\bibfnamefont {Rami}\
  \bibnamefont {Barends}}, \bibinfo {author} {\bibfnamefont {Julian}\
  \bibnamefont {Kelly}}, \bibinfo {author} {\bibfnamefont {Pedram}\
  \bibnamefont {Roushan}}, \bibinfo {author} {\bibfnamefont {Andrew}\
  \bibnamefont {Tranter}}, \bibinfo {author} {\bibfnamefont {Nan}\ \bibnamefont
  {Ding}},  \emph {et~al.},\ }\bibfield  {title} {\enquote {\bibinfo {title}
  {Scalable quantum simulation of molecular energies},}\ }\href@noop {}
  {\bibfield  {journal} {\bibinfo  {journal} {Physical Review X}\ }\textbf
  {\bibinfo {volume} {6}},\ \bibinfo {pages} {031007} (\bibinfo {year}
  {2016})}\BibitemShut {NoStop}%
\bibitem [{\citenamefont {Shen}\ \emph {et~al.}(2017)\citenamefont {Shen},
  \citenamefont {Zhang}, \citenamefont {Zhang}, \citenamefont {Zhang},
  \citenamefont {Yung},\ and\ \citenamefont {Kim}}]{shen2017quantum}%
  \BibitemOpen
  \bibfield  {author} {\bibinfo {author} {\bibfnamefont {Yangchao}\
  \bibnamefont {Shen}}, \bibinfo {author} {\bibfnamefont {Xiang}\ \bibnamefont
  {Zhang}}, \bibinfo {author} {\bibfnamefont {Shuaining}\ \bibnamefont
  {Zhang}}, \bibinfo {author} {\bibfnamefont {Jing-Ning}\ \bibnamefont
  {Zhang}}, \bibinfo {author} {\bibfnamefont {Man-Hong}\ \bibnamefont {Yung}},
  \ and\ \bibinfo {author} {\bibfnamefont {Kihwan}\ \bibnamefont {Kim}},\
  }\bibfield  {title} {\enquote {\bibinfo {title} {Quantum implementation of
  the unitary coupled cluster for simulating molecular electronic structure},}\
  }\href@noop {} {\bibfield  {journal} {\bibinfo  {journal} {Physical Review
  A}\ }\textbf {\bibinfo {volume} {95}},\ \bibinfo {pages} {020501} (\bibinfo
  {year} {2017})}\BibitemShut {NoStop}%
\bibitem [{\citenamefont {Kim}\ \emph {et~al.}(2020)\citenamefont {Kim},
  \citenamefont {Cao},\ and\ \citenamefont {Altman}}]{PhysRevB.101.125112}%
  \BibitemOpen
  \bibfield  {author} {\bibinfo {author} {\bibfnamefont {Jaewon}\ \bibnamefont
  {Kim}}, \bibinfo {author} {\bibfnamefont {Xiangyu}\ \bibnamefont {Cao}}, \
  and\ \bibinfo {author} {\bibfnamefont {Ehud}\ \bibnamefont {Altman}},\
  }\bibfield  {title} {\enquote {\bibinfo {title} {Low-rank sachdev-ye-kitaev
  models},}\ }\href {\doibase 10.1103/PhysRevB.101.125112} {\bibfield
  {journal} {\bibinfo  {journal} {Phys. Rev. B}\ }\textbf {\bibinfo {volume}
  {101}},\ \bibinfo {pages} {125112} (\bibinfo {year} {2020})}\BibitemShut
  {NoStop}%
\bibitem [{\citenamefont {Garc{\'\i}a-Garc{\'\i}a}\ \emph
  {et~al.}(2020)\citenamefont {Garc{\'\i}a-Garc{\'\i}a}, \citenamefont {Jia},
  \citenamefont {Rosa},\ and\ \citenamefont {Verbaarschot}}]{garcia2020sparse}%
  \BibitemOpen
  \bibfield  {author} {\bibinfo {author} {\bibfnamefont {Antonio~M}\
  \bibnamefont {Garc{\'\i}a-Garc{\'\i}a}}, \bibinfo {author} {\bibfnamefont
  {Yiyang}\ \bibnamefont {Jia}}, \bibinfo {author} {\bibfnamefont {Dario}\
  \bibnamefont {Rosa}}, \ and\ \bibinfo {author} {\bibfnamefont {Jacobus~JM}\
  \bibnamefont {Verbaarschot}},\ }\bibfield  {title} {\enquote {\bibinfo
  {title} {Sparse sachdev-ye-kitaev model, quantum chaos and gravity duals},}\
  }\href@noop {} {\bibfield  {journal} {\bibinfo  {journal} {arXiv preprint
  arXiv:2007.13837}\ } (\bibinfo {year} {2020})}\BibitemShut {NoStop}%
\bibitem [{\citenamefont {{Xu}}\ \emph {et~al.}(2020)\citenamefont {{Xu}},
  \citenamefont {{Susskind}}, \citenamefont {{Su}},\ and\ \citenamefont
  {{Swingle}}}]{2020arXiv200802303X}%
  \BibitemOpen
  \bibfield  {author} {\bibinfo {author} {\bibfnamefont {Shenglong}\
  \bibnamefont {{Xu}}}, \bibinfo {author} {\bibfnamefont {Leonard}\
  \bibnamefont {{Susskind}}}, \bibinfo {author} {\bibfnamefont {Yuan}\
  \bibnamefont {{Su}}}, \ and\ \bibinfo {author} {\bibfnamefont {Brian}\
  \bibnamefont {{Swingle}}},\ }\bibfield  {title} {\enquote {\bibinfo {title}
  {{A Sparse Model of Quantum Holography}},}\ }\href@noop {} {\bibfield
  {journal} {\bibinfo  {journal} {arXiv e-prints}\ ,\ \bibinfo {eid}
  {arXiv:2008.02303}} (\bibinfo {year} {2020})},\ \Eprint
  {http://arxiv.org/abs/2008.02303} {arXiv:2008.02303 [cond-mat.str-el]}
  \BibitemShut {NoStop}%
\bibitem [{\citenamefont {Crisanti}\ and\ \citenamefont
  {Sommers}(1992)}]{pspin}%
  \BibitemOpen
  \bibfield  {author} {\bibinfo {author} {\bibfnamefont {A.}~\bibnamefont
  {Crisanti}}\ and\ \bibinfo {author} {\bibfnamefont {H.~J.}\ \bibnamefont
  {Sommers}},\ }\bibfield  {title} {\enquote {\bibinfo {title} {The
  sphericalp-spin interaction spin glass model: the statics},}\ }\href
  {\doibase 10.1007/BF01309287} {\bibfield  {journal} {\bibinfo  {journal}
  {Zeitschrift f{\"u}r Physik B Condensed Matter}\ }\textbf {\bibinfo {volume}
  {87}},\ \bibinfo {pages} {341--354} (\bibinfo {year} {1992})}\BibitemShut
  {NoStop}%
\end{thebibliography}%

\end{document}